\documentclass[twocolumn,natbib,a4]{svjour3}         
\smartqed  
\usepackage{graphicx}
\usepackage{fancyhdr}

\bibpunct{[}{]}{,}{n}{}{;}

\usepackage{amsmath,wasysym,longtable,slashbox,txfonts,color,soul,fancyhdr}
\graphicspath{{figs/}}
\usepackage{epsfig,subfigure}
\usepackage{float}
\usepackage[misc]{ifsym}
\usepackage{subfigure}
\usepackage{pifont}

\usepackage{bm}        
\usepackage{amsfonts}
\usepackage{amssymb}

\newcommand*\diff{\mathop{}\!\mathrm{d}}

\definecolor{gray}{rgb}{0.8,0.8,0.8}
\DeclareSymbolFont{lettersA}{U}{txmia}{m}{it}
\DeclareMathSymbol{\muup}{\mathord}{lettersA}{22}
\DeclareMathSymbol{\piup}{\mathord}{lettersA}{25}

\def\footnoterule{}
\def\kh{\baselineskip}
\def\n{\noindent}
\def\[2mm]{\cr\noalign{\vskip2mm}}

\newcommand{\rb}{Rayleigh-B\'{e}nard~}
\newcommand{\pb}{~Prandtl-Blasius-Pohlhausen~}

\def\n{\noindent}\def\fo{\footnotesize}

\makeatletter
\def\footnoterule{\relax%
	\kern-5pt
	\hbox to \columnwidth{\hfill\vrule width 0.99\columnwidth height 0.4pt\hfill}
	\kern4.6pt}
\makeatother

\makeatletter
\renewcommand{\@thesubfigure}{\hskip\subfiglabelskip}
\makeatother

\begin{document}
\def\makeheadbox{\relax}
\renewcommand{\headrulewidth}{0pt}
    
\title{Boundary layer structure in turbulent \rb convection in a slim box
	\thanks{Grants or other notes about the article that should go on the front page should be placed here. General acknowledgments should be placed at the end of the article.}
}

\titlerunning{Boundary layer structure in turbulent \rb convection in a slim box}        

\author{Hong-Yue Zou$^1$         \and
        Wen-Feng Zhou$^1$ \and Xi Chen$^1$         \and
        Yun Bao$^2$ \and Jun Chen$^{1,\ast}$         \and
        Zhen-Su She$^1$ 
}

\authorrunning{H.-Y. Zou, W.-F. Zhou et al.} 

\institute{J. Chen \at
              Peking University \\
              Tel.: +86-10-62756559\\
              \email{jun@pku.edu.cn}           
           \and
}

\date{Received: date  / Accepted: date}


\twocolumn[
\setcounter{page}{1} 
\fancyhead{} 
\vspace*{-3cm}



\vspace*{-3mm}


\vskip -0.55mm \n{\colorbox{gray}{ \vphantom{ $ \big| $ }RESEARCH
		\ PAPER \qquad \qquad \qquad \qquad \qquad \qquad \ }
	
	\label{first-page}
	
	\vskip 1cm}

\maketitle
\begin{@twocolumnfalse}
\abstract{The logarithmic law of mean temperature profile has been observed in different regions in \rb turbulence. However, how thermal plumes correlate to the log law of temperature and how the velocity profile changes with pressure gradient are not fully understood. Here, we performed three-dimensional simulations of \rb turbulence in a slim-box without the front and back walls with aspect ratio, $\mathrm{width}:\mathrm{depth}:\mathrm{height}=L:D:H=1:1/6:1$ (respectively corresponding to $x$, $y$ and $z$ coordinates), in the Rayleigh number $Ra = [1\times10^8, 1\times10^{10}]$ for Prandtl number $Pr=0.7$. To investigate the structures of the viscous and thermal boundary layers, we examined the velocity profiles in the streamwise and vertical directions (i.e. $U$ and $W$) along with the mean temperature profile throughout the plume-impacting, plume-ejecting, and wind-shearing regions. The velocity profile is successfully quantified by a two-layer function of a stress length, $\ell_u^+\approx \ell_0^+(z^+)^{3/2}\left[1+\left({z^+}/{z_{sub}^+}\right)^4\right]^{1/4}$, as proposed by She et al. \textbf{(She 2017)}, though neither a \pb type nor the log-law is seen in the viscous boundary layer. In contrast, the temperature profile in the plume-ejecting region is logarithmic for all simulated cases, being attributed to the emission of thermal plumes. The coefficient of the temperature log-law, $A$ can be described by composition of the thermal stress length $\ell^*_{\theta 0}$ and the thicknesses of thermal boundary layer $z^*_{sub}$ and $z^*_{buf}$, i.e. $A \simeq z^*_{sub}/\left(\ell^*_{\theta 0}{z^*_{buf}}^{3/2}\right)$. The adverse pressure gradient responsible for turning the wind direction contributes to thermal plumes gathering at the ejecting region and thus the log-law of temperature profile. The Nusselt number scaling and local heat flux of the present simulations are consistent with previous results in confined cells. Therefore, the slim-box RBC is a preferable system for investigating in-box kinetic and thermal structures of turbulent convection with the large-scale circulation on a fixed plane.
}

	\vskip 3mm
\end{@twocolumnfalse}
]
\vskip 3mm

{\footnotetext[1]{\vspace{0.05em}
\hspace{-2em} \noindent Corresponding author:~Jun Chen \\
			jun@pku.edu.cn\\
			\vspace{0.1em}\\
			$^1$ State Key Laboratory for Turbulence and Complex Systems, Department of Mechanics and Engineering Science, College of Engineering, Peking University, Beijing 100871, China\\
			\vspace{0.1em}\\			$^2$ Department of Mechanics, College of Engineering, Sun Yat-sen University, Guangzhou 510275, China
}
}


\section{Introduction}\label{sec:intro}
Rayleigh-B{\'e}nard convection (RBC) is commonly used to study natural convection due to the simplicity of its configuration and the richness of its flow regimes. In this system, fluid is filled in a closed cell, heated on the bottom and cooled on the top, with adiabatic side no-slip wall \cite{Xia2013TAML, Liu2017AMS, Gao2018CPB}. The control parameters are the Rayleigh number $Ra=g\beta \Delta H^3/(\nu\kappa)$, the Prandtl number, $Pr=\nu/\kappa$, and the aspect ratio $\Gamma = L/H$, where $\nu$ is the kinematic viscosity, $\kappa$ the thermal diffusivity, $H$ the height of the sample, $L$ its width, $g$ the gravitational acceleration, and $\beta$ the thermal expansion coefficient, respectively. Enhancement of the heat transport of a natural convection system, like the RBC, is particularly useful in many industrial processes and is of fundamental interest \cite{Chen2017IJHMT, Bao2015JFM}.

The boundary layer (BL) in RBC exhibits a transition from laminar to turbulent regime when $Ra$ exceeds a critical value, $Ra_c$. In the laminar regime, the mean velocity profile (MVP) takes a \pb (PBP) type, whereas in turbulent regime, usually a logarithmic (log) profile is expected, when an analogy is made with in a BL passing a flat plate. Recently, a BL equation for RBC of $Pr>1$ has been developed \cite{Shishkina2015PRL}, considering both laminar and turbulent contributions. Lately, a model describing the mean profiles of temperature and its variance in the near-wall region was reported and experimentally tested \cite{Wang2016PRF}. For $Ra<Ra_c$, the RBC-BL presents a PBP-like profile when measured at near-wall regions in zero pressure gradient (ZPG) condition \cite{Zhou2010JFM, vanderPoel2013JFM}. Nevertheless, the lateral change of the profile is remarkable, presenting significant deviations from the PBP-type at high $Ra$ number \cite{Wagner2012JFM, Shi2012JFM}.

The logarithmic mean temperature profile (MTP) is another issue which has recently caught great attention in numerical and experimental studies, for $Ra$ ranging from $10^{10}$ to $10^{15}$ \cite{Ahlers2014JFM, Ahlers2012PRL, Wei2014JFM}. Logarithmic temperature profiles were observed near the sidewalls in experiments and DNS, and its thickness becomes remarkably decreases near the middle of the conducting plate. The mechanism behind the log-law of the MTP is still in debate --- whether it is induced by the momentum transport near the no-slip sidewall or by the heat transport due to emitted plumes or by both is still unknown. Some results indicate the correlation between the intensive plume emission and the log-law of the temperature profile. In a cylindrical containers at $\Gamma\simeq 1$, the plumes are found to be abundantly emitted from the top/bottom plate near the sidewalls, leading to an intensive local heat flux \cite{Shishkina2007PoF}. In a recent two-dimensional (2D) DNS study with horizontal periodic boundary, it is seen that a vertical log temperature profile only appears in the regions where plumes accumulate \cite{vanderPoel2015PRL}. However, the physics in these regions is still not fully understood.

Most previous experimental and numerical studies of the RBC system set the horizontal section in a circular or square geometry, where the LSC exhibits frequent reversal, cessation or azimuthal motion \cite{Huang2015PRL, Xi2016JFM}. In such systems, the plume-emitting regions appear to `wander' along the conducting plate, and it is hard to extract the property of a certain flow region free of the influence from other regions. Similar phenomena have been observed in the cubic box ($L:D:H=1:1:1$) where the large-scale convection is found to exhibit random reorientation of LSC and low-frequency oscillation perpendicular to LSC \cite{Vasiliev2016IJHMT}. Complex convection flow also appears in a rectangular container for $Ra=8\times 10^8 \sim 1\times10^{10}$, due to strong secondary flow in the form of horizontal rolls surrounding the core of the cell and orthogonal to the cross-stream rolls \cite{Podvin2012PoF}. Thus, neither a cubic cell nor a rectangular cavity is appropriate to establish a turbulent RBC with a statistical steady LSC over a large range of $Ra$.

We intend to perform a 3D simulation with an LSC on a fix plane so that a statistical mean field can be studied in great detail. This is achieved by reducing the scale in the depth direction to make a slim-box, e.g. $L:D:H=1:1/6:1$, for which the mean flow becomes ideally confined on the vertical plane. It is reported that the aspect ratio in the planes perpendicular to the LSC, $D/L$, has strong effects on the global heat transport --- The increased wall friction and suppressed LSC lead to more coherent and energetic plumes emitting from the conducting plates and thus enhancement of the global heat transport \cite{Huang2013PRL}. The focus of the present study is to investigate a steady LSC on a plane free from the wall effect, rather than the confinement effect. This helps to settle down a better defined statistical mean field with well-defined flow regions. The periodic condition in the depth ($y$-)direction allows 3D fluctuations of both velocity and temperature over a range of scales smaller than $1/6$ of the depth of the box, which are wide enough to develop relevant turbulent thermal convection for the simulated $Ra$. This configuration will be referred to as the slim-box RBC simulation, which has a depth extent large enough for developing 3D fluctuations, and slim enough to confine the LSC on a plane (i.e. $D/L=1/6$). Compared to other RBC system of $D/L\simeq 1$, the slim-box simulation establishes a stronger and more stable LSC with fixed wind direction. It will be seen that the periodic boundary condition in $y$ somewhat mimics the cylinder cell; indeed, the measured velocity and temperature profiles averaged in the depth direction present relevant features of experimental and numerical results previously observed.

We focus on the characteristics of the convection flow in different regions. An outcome of the current simulation is, in addition to LSC and the corner roll \cite{Zhou2018PoF}, the identification of three flow regions, namely impacting, wind-shearing, and ejecting regions in time-averaged velocity field, similar to the results in 3D DNS of the RBC in a circular cylinder \cite{vanderPoel2015JFM}. Note that the RBC in slim-box has a larger LSC, a result of the absence of friction due to sidewalls. The coherent motion of the stronger LSC yields a thinner viscous BL and hence a larger Reynolds number. The Nusselt number $Nu$ in the present study is also slightly larger than that in a confined cell, as measured at the same $Pr$ \cite{Chong2016JFM}.

The mean horizontal and vertical mean velocity ($U$ and $W$) and temperature profiles in the slim-box (averaged in the depth direction) are measured and studied in great detail in three regions, e.g. ejecting, wind-shearing, and impacting regions, for the medium Rayleigh number. Due to the strong adverse pressure gradient in the wind-shearing and ejecting regions, the BL in RBC is very different from a turbulent BL on a flat plate, so that neither a PBP-type nor the log-law BL is indeed observed in the MVP. On the other hand, the multi-layer theory proposed by She et al. \cite{She2017JFM} allows us to analytically quantify the kinetic BL by means of a two-layer stress length function. 

For the MTP, the ejecting region is found to hold a logarithmic region in all simulated cases, in agreement with previous studies. This logarithmic layer is, however, one of the multi-layer structure for the temperature length, and the log-law coefficient, $A$, can be related to the thickness of sub and buffer layer, which are measured at all streamwise location and shown to exhibit also a two-layer structure away from the ejection corner of the RBC cell. This model yields a two-dimensional temperature mean-field, in agreement with DNS data, for the first time, which is superior to the result of Grossmann et al. \cite{Grossmann2012PoF} (i.e. $|A|=|A_1|/\sqrt{4x(1-x)}$). The relation between the log law of MTP and the emission of thermal plumes, previously suggested by a two-dimensional simulation of RBC  \cite{vanderPoel2015PRL}, is confirmed in our 3D simulations, and the role of the adverse pressure gradient is emphasized. 

The paper is organized as following. In \S2, we discuss the numerical simulation method. \S3 contains the results and discussions, including temperature and pressure distribution, the man velocity and temperature profiles, and the heat flux distribution in the slim-box. \S4 is the conclusion.

\section{Numerical Setup}
\begin{figure}
\includegraphics[width=0.46\textwidth]{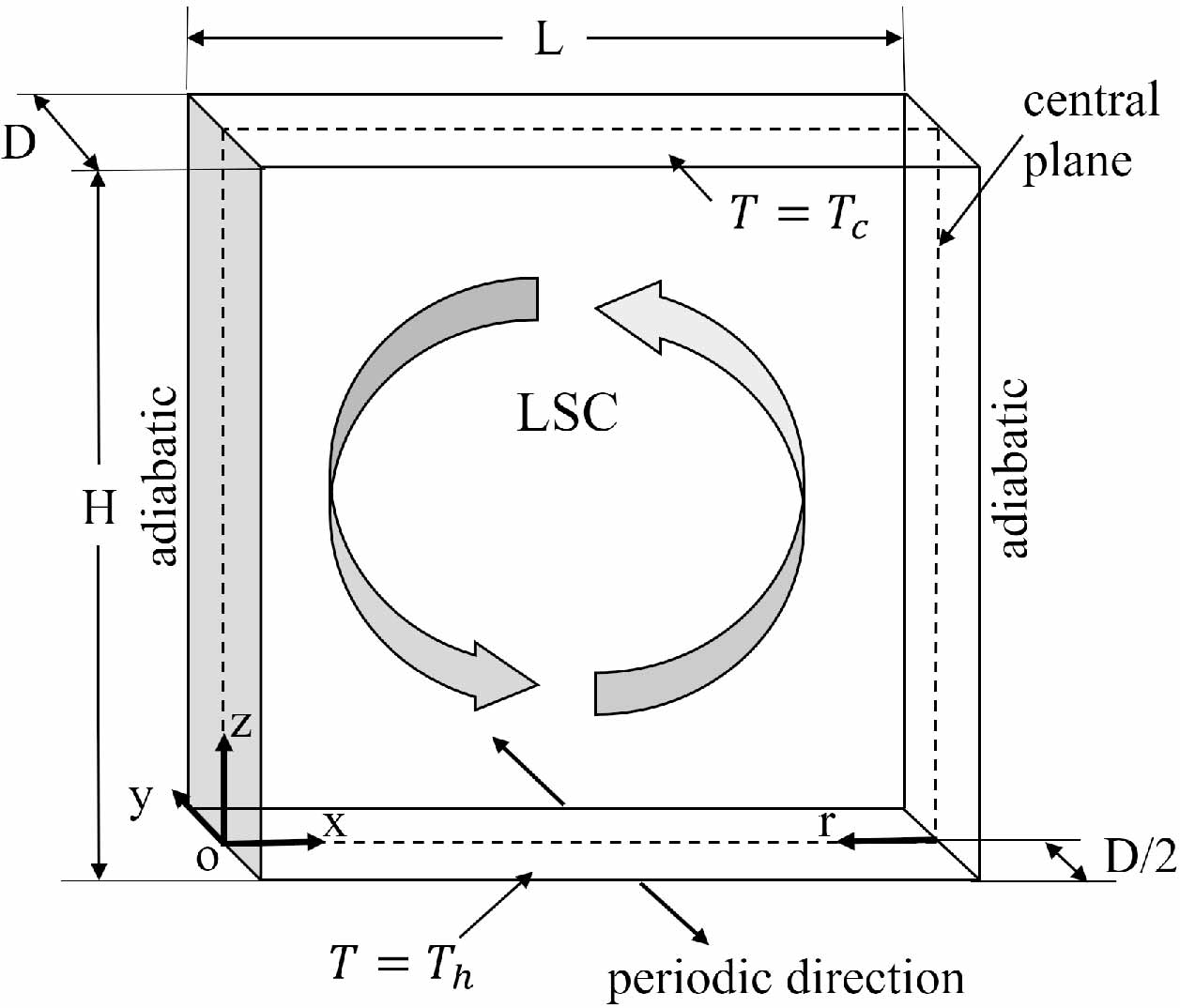}
\caption{Sketch of the convection box showing the definitions of coordinates and dimensions. The plane marked by dashed-line is the location of the side-view of the convection. \label{fig:schm}}
\end{figure}

The choice of the geometrical configuration is based on the following considerations. When the scale in the depth direction is reduced to form a slim-box (with, e.g. $L:D:H=1:1/6:1$, where $L$, $D$, $H$ denote horizontal $x$, depth $y$ and vertical $z$ directions, respectively, and with a vertical aspect ratio $\Gamma=L/H=1$), the mean flow is ideally confined within the vertical plane (at least approximately), see Fig. \ref{fig:schm} which displays the geometry and the LSC schematically. The periodic boundary condition set in the $y$ direction allows 3D fluctuations of both velocity and temperature over a range of scales smaller than $1/6$ of the length of the box, in which the flow becomes turbulent.

We numerically integrate the incompressible Navier-Stokes equation with Boussinesq's approximation and the continuity equation, following ref. \cite{Verzicco2003JFM}:
\begin{eqnarray}\label{eq:NS}
\frac{\partial \overrightarrow{u}}{\partial t} +(\overrightarrow{u}\cdot \nabla)\overrightarrow{u}=-\nabla p +\theta \overrightarrow{z} +\nu\nabla^2\overrightarrow{u};\\\label{eq:T}
\frac{\partial \theta}{\partial t} + (\overrightarrow{u}\cdot \nabla) \theta = \kappa \nabla^2 \theta;\\\label{eq:mass}
\nabla\cdot\overrightarrow{u} = 0,
\end{eqnarray}
\noindent where $\overrightarrow{z}$ is the unity vector pointing in the opposite direction to gravity, $\overrightarrow{u}$ the velocity vector, $p$ the pressure and $\theta$ the non-dimensional temperature (with $\pm1/2$ at the bottom and top walls), respectively. The integrated equations are normalized using the free-fall velocity $U=\sqrt{RaPr}(\kappa/H)$, the temperature difference between the upper and lower conducting plates $\Delta$(=1), the pressure $P_0=RaPr(\rho\kappa^2)/H^2$, and the time scale $T_0=(H^2/\kappa)\sqrt{RaPr}$.

The fluid in the slim-box RBC is bounded in the $x-z$ plane by the upper- and lower-isothermal plates and adiabatic sidewalls, so the boundary conditions are $\partial{\theta}/\partial{x}|_{x=0}=\partial{\theta}/\partial{x}|_{x=L}=0$. No-slip and impenetrability conditions are used for all solid boundaries. Periodicity is assumed in the $y$ direction: i.e. $\theta|_{y=0}=\theta|_{y=D}$, $\partial{\theta}/\partial{y}|_{y=0}=\partial{\theta}/\partial{y}|_{y=D}$ and $\overrightarrow{u}|_{y=0}=\overrightarrow{u}|_{y=D}$, $\partial{\overrightarrow{u}}/\partial{y}|_{y=0}=\partial{\overrightarrow{u}}/\partial{y}|_{y=D}$).

All the simulations have been performed with a second-order finite-difference code, see ref. \cite{Verzicco1996JCP} for details. Due to the absence of singularity at the origin, we use the central second-order finite difference in the $y$ direction as well. The time-advancement applies a time-splitting method which has been extensively discussed in ref. \cite{Kim1985JCP,Rai1991JCP}. The third-order low-storage Runge--Kutta in conjunction with the Crank--Nicolson scheme is applied to evaluate the nonsolenoidal velocity \cite{Rai1991JCP}. The finite-difference scheme for the temperature equation is the same as velocity, except for pressure-related terms.

Solving the Poisson equation for pressure requires that the solution be sufficiently smooth up to the boundary. Iterative method is found inefficient at high $Ra$ when small-scale fluctuations are abundantly developed. We thus applied the FFT method to reduce the PDD solver from 3D to 2D; see \cite{Sun1995PC} for details.

Keeping the grid spacing smaller than the Kolmogorov scale $\eta_K$ and the Batchelor scale $\eta_B$ over the whole domain is important to ensure proper spatial resolution \cite{Shishkina2010NJP}, where $\eta_K=(\nu^3/\epsilon_u)^{1/4} = \frac{Pr^{1/2}}{Ra^{1/4}(Nu-1)^{1/4}}H$, and $\eta_B = \eta_K/\sqrt{Pr}$ \cite{Sheel2013NJP}. The time step is chosen to satisfy the Courant--Friedrichs--Lewy (CFL) condition, i.e. $CFL \leq 0.2$. In our simulations, $Nu$ was calculated by integrating over the whole volume and over time.  Table \ref{Table-I} reports the minimum and maximum grid spacings, $\Delta_{min}$ and $\Delta_{max}$, which are indeed smaller than $\eta_{B}$ and $\eta_K$; thus, the finest scales in the bulk flow are well resolved. Then, the thermal BL thickness is calculated using the relation of $\lambda_\theta=H/2Nu$. The Bolgiano length scale is evaluated using  $L_{B}={\langle\epsilon_u\rangle^{5/4}}/{(g^2\beta^2\langle\epsilon_{\theta}\rangle)^{3/4}}\approx\frac{Nu^{1/2}}{(Pr\cdot Ra)^{1/4}}H$, where $\epsilon_{\theta}$ is the thermal dissipation rate \cite{Shishkina2010NJP}.

\begin{table*}
\tabcolsep 2.mm
\caption{The numbers of the grids $N_x\times N_y \times N_z$,values of the Nusselt number estimated by three different methods $Nu_1$, $Nu_2$, $Nu_3$, the minimum and maximum slim-box sizes  $\Delta_{min}$ and $\Delta_{max}$, the Bolgiano length scale $L_{B}$, the Kolmogorov viscous scale $\eta_K$, and the Batchelor scale $\eta_{B}$. \label{Table-I}}
\begin{center}
{\scriptsize
	\begin{tabular}{lccccccccc}
	\hline
	\hline
	$Ra$ & $N_x\times N_y \times N_z$ & $Nu_1$& $Nu_2$ & $Nu_3$ & $\Delta_{min}$ & $\Delta_{max}$  &$L_{B}$  & $\eta_K$ & $\eta_{B}$\\
	\hline
	$1\times 10^8$ & $768\times 128\times 512$   &34.7 & 34.2 & 34.3 & $0.23\times10^{-3}$ & $4.2\times10^{-3}$   & $6.44\times10^{-2}$  & $3.47\times10^{-3}$ & $4.15\times10^{-3}$\\
	$5\times 10^8$ & $768\times 128\times 512$   &55.8 & 54.8 & 55.3 & $0.45\times10^{-3}$ & $1.4\times10^{-3}$   & $5.42\times10^{-2}$  & $2.00\times10^{-3}$ & $2.41\times10^{-3}$\\
	$1\times 10^9$ & $768\times 256\times 800$   &68.9 & 69.3 & 68.5 & $0.08\times10^{-3}$ & $1.6\times10^{-3}$   & $5.10\times10^{-2}$  & $1.64\times10^{-3}$ & $1.96\times10^{-3}$\\
	$5\times 10^9$ & $1024\times 256 \times 800$   &110.8 & 109.1 & 109.7 & $0.20\times10^{-3}$ & $1.8\times10^{-3}$   & $4.31\times10^{-2}$  & $0.97\times10^{-3}$ & $1.20\times10^{-3}$\\
	$1\times 10^{10}$ & $1600\times 512\times 1600$  &144.6 & 147.4 & 143.2 & $0.05\times10^{-3}$ & $0.8\times10^{-3}$  & $4.16\times10^{-2}$  & $7.60\times10^{-4}$ & $9.08\times10^{-4}$\\
	\hline
	\hline
	\end{tabular}
}
\end{center}
\end{table*}

An additional control of the DNS quality is to compare different $Nu$ by different integration procedures. One procedure is to compute directly the heat flux by integrating along the two conducting walls, i.e. $Nu_1=-\partial{\Theta}/\partial{z}$ at the upper ($z=H$) or the below ($z=0$) plates. The second procedure is to compute the volume-averaged temperature dissipation $\varepsilon_\theta$ or energy dissipation $\varepsilon_u$ \cite{Shraiman1990PRA}, yielding  $Nu_2=\langle\varepsilon_\theta\rangle$ or $Nu_3=1+Pr \langle\varepsilon_u\rangle$, respectively, where $\langle \cdot \rangle$ denotes the ensemble average over both time and space. These Nusselt numbers are calculated and reported also in Table \ref{Table-I}, in good agreement with each other, ensuring the statistical convergence of the simulated RB system.

\begin{figure}
\includegraphics[width=0.5\textwidth]{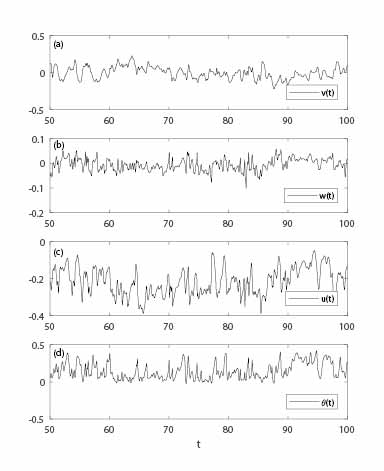}
\caption{Signals of a numerical probe located halfway down the slim-box at $z/H=9.8\times10^{-3}$, $x/L = 0.5$ and $y/D = 0.5$, for $Ra=1\times10^8$. \label{fig:signal}}
\end{figure}

A well-resolved simulation at high $Ra$ requires a large computation resource. The simulations employed up to $1024$ TH-1A CPUs (central processing units), using $500\times128\times768$ grids for $Ra=1\times10^8$ and $10^9$, and $1600\times512\times1600$ grids for $Ra=1\times10^{10}$. 
In order to examine the turbulent state of the RBC flow, we placed $8$ probes at different heights, from $z/H=9.8\times10^{-3}$ to $z/H=0.25$ at the central horizontal and depth location, namely, $x/L=\frac{1}{2}$ and $y/D=\frac{1}{2}$. These probes record point-wise fluctuating temperature $\theta(t)$ and velocity (of three components $u(t)$, $v(t)$, and $w(t)$). Fig. \ref{fig:signal} shows that the flow is in a developing stage before $30$ dimensionless time, and then reaches a statistically steady state around $t=40$, when we begin to carry out the average calculation. It is also seen that the positive correlation between velocity along the LSC and the temperature fluctuation, as respectively seen in Fig. \ref{fig:signal}(c) and (d).

\section{Results and Discussion}
\subsection{Temperature and pressure distributions}
\begin{figure*}
\subfigure[\label{subfig:t-3d-R8}]{\includegraphics[width = .32\textwidth]{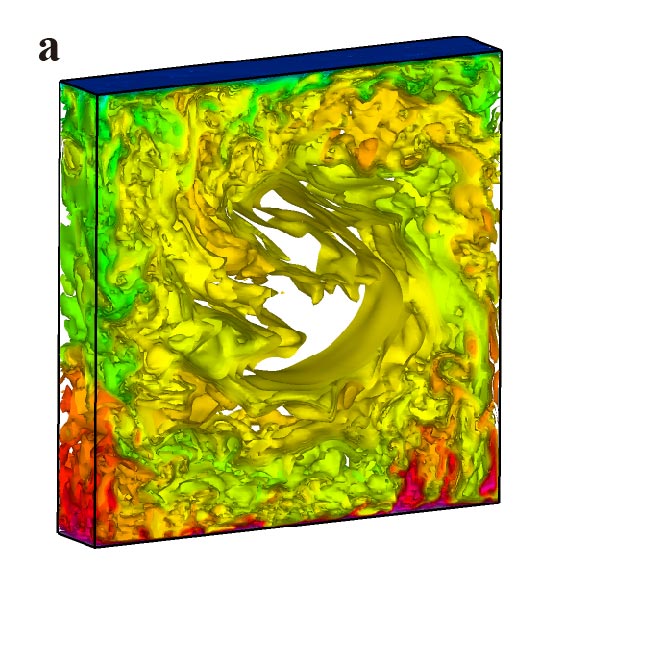}}
\subfigure[\label{subfig:t-slice-R8}]{\includegraphics[width = .32\textwidth]{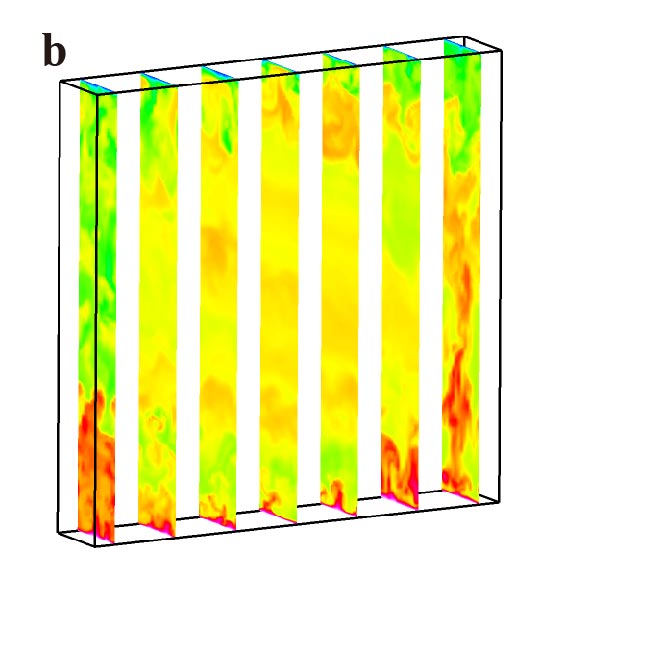}}
\subfigure[\label{subfig:t-2d-R8}]{\includegraphics[width = .32\textwidth]{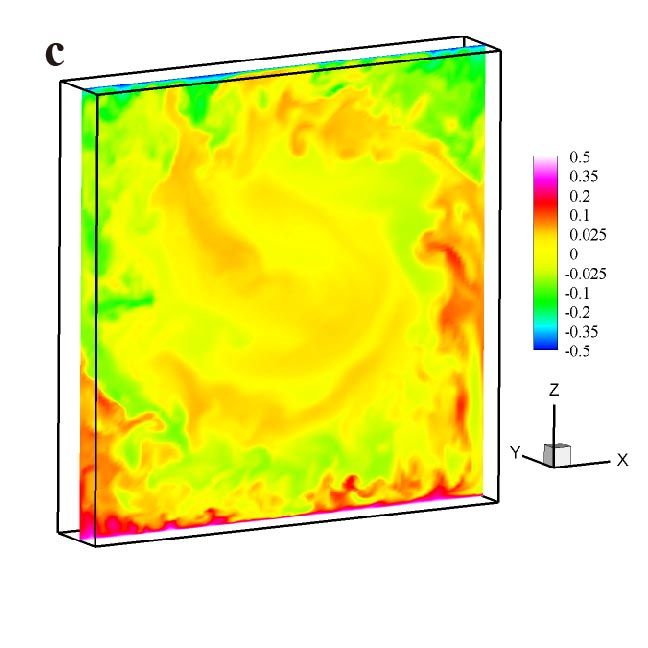}}
\subfigure[\label{subfig:p-3d-R8}]{\includegraphics[width = .32\textwidth]{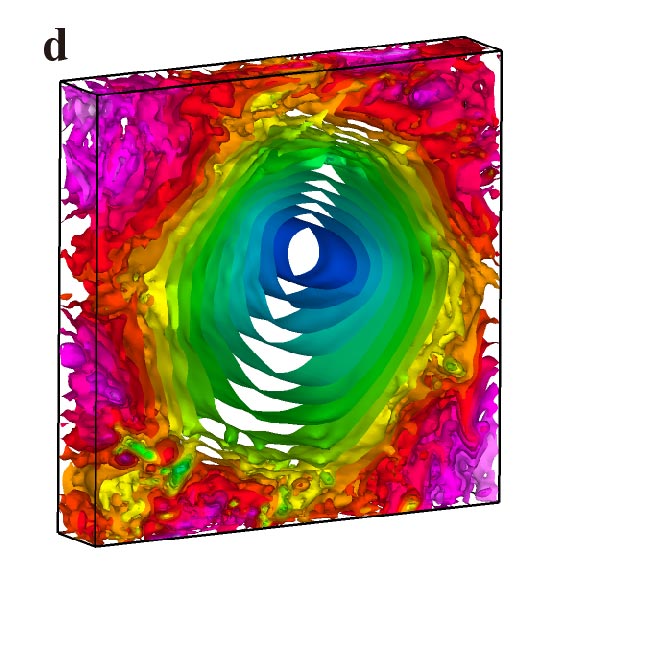}}
\subfigure[\label{subfig:p-slice-R8}]{\includegraphics[width = .32\textwidth]{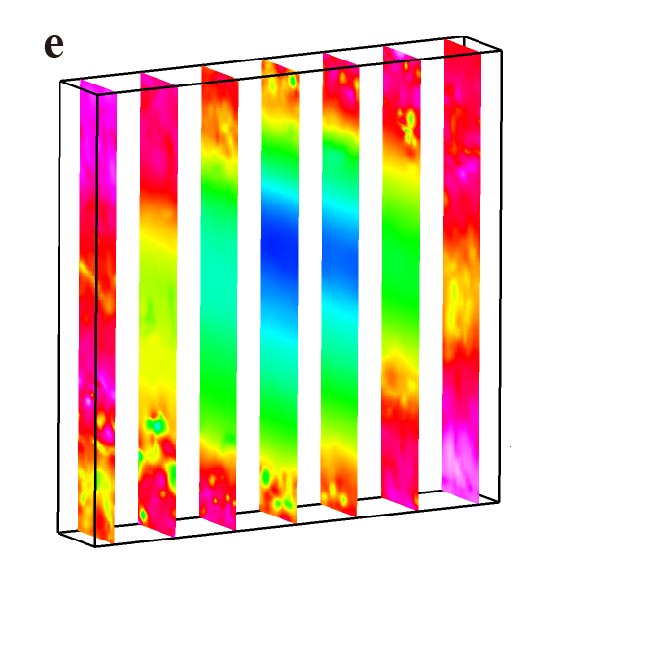}}
\subfigure[\label{subfig:p-2d-R8}]{\includegraphics[width = .32\textwidth]{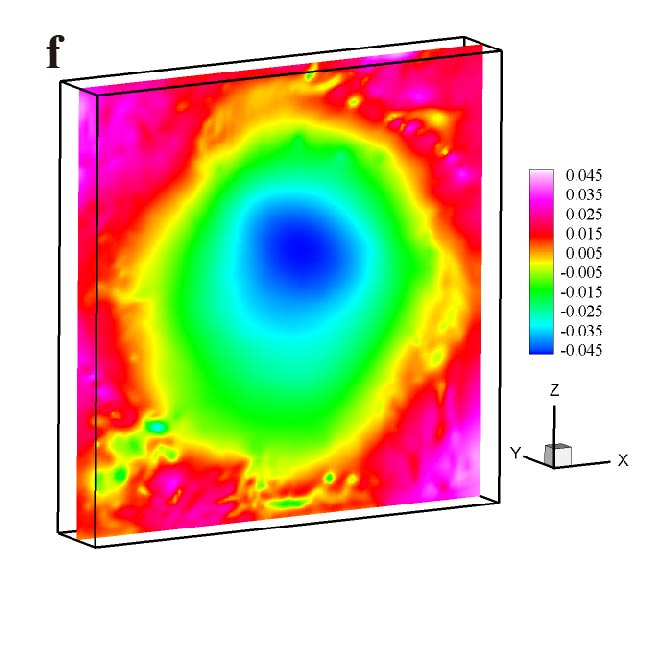}}
\caption{(Color online) Visualization of the instantaneous temperature (a)--(c) and pressure (d) -- (f): The color in (a) -- (c) denotes temperature, and the color in (d) -- (f) denotes pressure. The slices in (b) and (e) are the corresponding fields on the $yz$-planes at various streamwise ($x$) locations. The fields shown in (c) and (f) are the fields on the $xz$-plane on the halfway on the depth of the box. These snapshots are obtained from the simulation at $Ra=1\times10^8$ and $Pr=0.7$.}
\end{figure*}

Fig. \ref{subfig:t-3d-R8} and \ref{subfig:p-3d-R8} show that the LSC is confined on the $x-z$ plane, presented by the snapshots of temperature and pressure isosurface for $Ra=1\times 10^8$. Three-dimensionality is remarkable in the near-wall region, where thermal plumes are intensively emitted. However, in LSC, the variances of temperature and pressure in $y$ direction are rather weak, indicating that a quasi-2D flow at the center of the box is dominant in the slim-box simulation.

The data collected for comparison with experimental and other DNS studies include the viscous BL thickness and heat flux. Two sets of experimental measurements \cite{Sun2008JFM, Wei2013JFM} and two sets of DNS from \cite{Kaczorowski2014JFM} with different aspect ratios, and from Stevens et al. \cite{Stevens2011JFM} for a cylindrical box with aspect ratio $\Gamma=1/2$ at higher $Ra$ of $2 \times 10^{10}$, $2 \times 10^{11}$ and $2 \times 10^{12}$. Streamwise mean velocity measurements from du Puits et al. \cite{duPuits2007PRL} at several $Ra$s around $10^{11}$ and mean temperature measurements from Ahlers et al. \cite{Ahlers2012PRL} are also considered. For convenience, we summarize these DNS and experimental results in Table \ref{tab:paras}.
\begin{table*}
\tabcolsep 1.mm
\caption{The parameters of DNS and experiments.\label{tab:paras}}
\begin{center}
{\scriptsize
	\begin{tabular}{clccccc}
	\hline
	\hline
	$Case$  & Reference &   $Ra$  &    $Pr$    & $L:D:H$ (rectangular) or $D:H$ (cylindrical) &  Confinement shape     &   $Data$ \\
	\hline
	$DNS-A$ & Present study& $1\times10^8 \sim 1\times 10^{10}$  &    0.7   & $1:1/6:1$ & rectangular\footnote{a slim-box with the periodic condition in $y$ direction} &   $U,W,\theta,Nu,\lambda_u$      \\
	$DNS-B$ & \cite{Stevens2011JFM}  &$2\times10^{10} \sim 2\times10^{12}$   &    0.7  & $1:2$ &  cylindrical    &   $Nu,\theta$  \\
	$DNS-C$ & \cite{Kaczorowski2014JFM} & $1\times10^7 \sim 1\times 10^{10}$  &    $0.7$, $4.38$    & $1:1/64:1\sim 1:1:1$  &  rectangular    &   $Nu$ \\
	$EXP-A$ &  \cite{duPuits2007PRL} & $1.2\times10^{11} \sim 9.8\times10^{11}$         &    0.7         & $1:1.13$ &  cylindrical    &   $U$  \\
	$EXP-B$ & \cite{Ahlers2012PRL} & $4\times10^{12}\sim 1\times10^{15}$   &    0.8  & $1:2$ &  cylindrical &  $\theta$   \\
	$EXP-C$ & \cite{Sun2008JFM} &$1\times10^{9} \sim 1\times10^{10}$              &    4.3  & $25:7:24$          &  rectangular    &   $Re,\lambda_u,U$       \\
	$EXP-D$ & \cite{Wei2013JFM} & $2.4\times10^{8}\sim 5.6\times10^9$  & 5.3  & $1:1$ &  cylinder  &   $Re,\lambda_u$   \\
	\hline
	\hline
	\end{tabular}
}	\end{center}
\end{table*}

\begin{figure*}
\begin{center}
\subfigure[\label{subfig:Figure4a}]{\includegraphics[width = 0.45\linewidth]{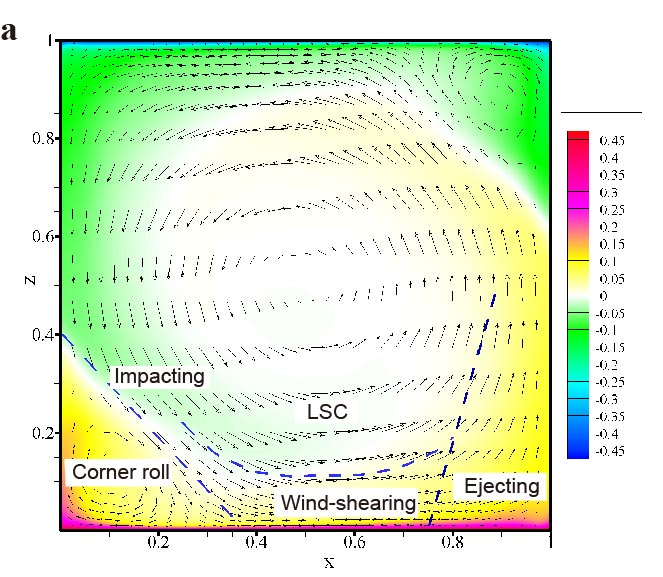}}
\subfigure[\label{subfig:Figure4b}]{\includegraphics[width =  0.45\linewidth]{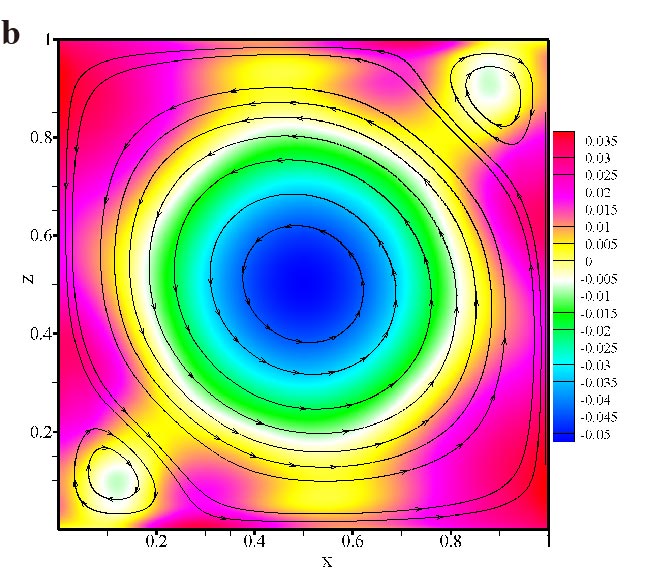}}
\subfigure[\label{subfig:Figure4c}]{\includegraphics[width =  0.45\linewidth]{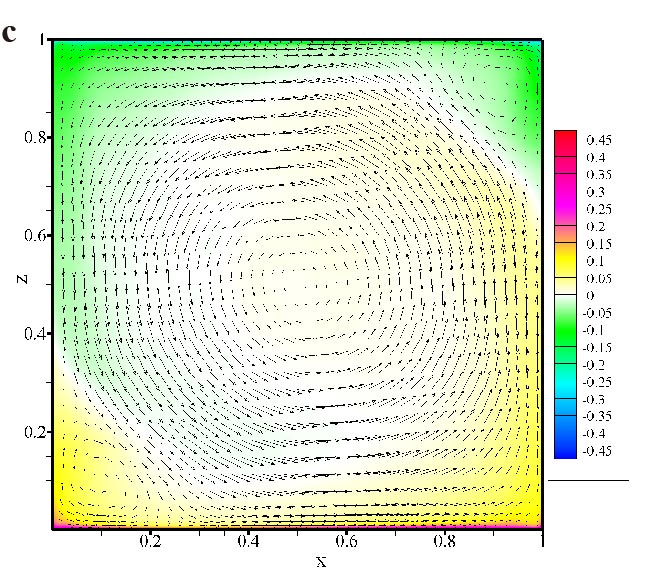}}
\subfigure[\label{subfig:Figure4d}]{\includegraphics[width =  0.45\linewidth]{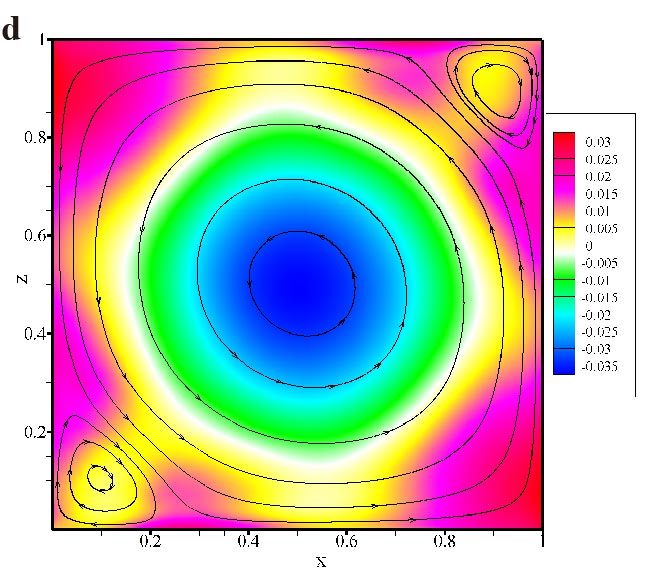}}
\end{center}
\caption{(Color online) The time-averaged 2-D distributions of pressure and temperature displayed by pseudo-colors at $Ra = 1\times10^8$ in (a) -- (b) and $Ra = 1\times10^9$ in (c) -- (d), respectively. The arrows indicate the velocity vectors. The rotatory LSC, the corner roll, the impacting region, the wind-shearing region and the ejecting region are marked by dashed lines in (a). The lines with arrows in (b) and (d) are the time-averaged streamlines. \label{fig:TP-field}}
\end{figure*}

Ensemble average is carried out by integrating in the depth ($y$) direction and in time. The mean velocity, temperature and pressure distributions at $Ra=1\times 10^8$ and $1\times 10^9$ are shown in Fig. \ref{fig:TP-field}. The mean flow is represented by a counter-clockwise rotation at the center of the box. As $Ra$ increases, the high speed regions are shifted towards the perimeter of the box with a relatively more slowly moving central region. It is noted that two pairs of corner rolls are observed, which are considered as the secondary flow induced by the LSC, as being reported previously in experimental \cite{Krishnamurti1981ProcNAS, Qiu1998PREb, Niemela2001JFM} and numerical studies \cite{Benzi2008EPL, Sugiyama2009JFM, Shi2012JFM}. Particularly, a model for the corner rolls in an RB cell was suggested by Zhou et al. \cite{Zhou2018PoF}. 

Fig. \ref{fig:TP-field} shows that the positive high pressure coincides with the regions of intensively ejecting plumes, corresponding to the lower right and upper left corner of the slim-box. On the other hand, the center of the box or of LSC corresponds to the low-pressure zone, as shown in Fig. \ref{subfig:Figure4b} and \ref{subfig:Figure4d}. This indicates that pressure gradient maintains a balance to the centrifugal force. Similarly, the cores of two corner rolls are closely associated with two lower pressure zones, which was also observed in 2D simulations \cite{Zhou2011PoF}. It is seen that the 3D corner roll is a bit smaller than the 2D one at $Ra=1\times10^8$. Furthermore, the 3D simulation presents a more round shape of LSC and a perceivably wider wind-shearing region than the 2D simulation, indicating that the 3-dimensionality of the near-wall corner flow is remarkable. This observation is consistent with the results of the previous numerical study \cite{Zhou2018PoF}.

Based on the observation of the averaged flow on the $xz$-plane, we characterize the RBC flow by five regions: (a) the LSC motion in bulk, (b) the corner roll, (c) the impacting regions above the corner rolls with ($0 \leq x \leq 0.25$), (d) the wind-shearing regions towards the middle of the conducting plates ($0.25 \leq x \leq 0.75 $), and (e) the ejecting region ($0.75 \leq x \leq 1 $). The last three regions clearly possess a turbulent BL at large aspect ratio, cf. \cite{vanderPoel2015PRL}. Note that the change of flow direction at the corners signifies the presence of high pressure gradient \cite{vanReeuwijk2008PREa, vanReeuwijk2008PREb}, as consistently presented in Fig. \ref{subfig:Figure4c} and \ref{subfig:Figure4d}.

According to the correlation between pressure and the velocity fields, one can see that, in the wind-shearing region, the BL flow is driven by a pressure gradient. The fluid near the bottom wall advects from a favorable pressure gradient region to an adverse pressure gradient (APG) region. The APG in respect to the flow direction is present in the ejecting region, and responsible for the change of the flow from horizontal to vertical direction. This makes RBC-BL flow distinct from canonical BL under zero-pressure-gradient condition. Consequently, the BL of a turbulent RBC is inevitably of non-PBP-type. In \S \ref{subsec:viscBL}, we will discuss the structure of the BL by analyzing the MVP in detail.

\begin{figure*}
\begin{center}
\subfigure[\label{subfig:Figure5a}]{\includegraphics[width =  0.21\textwidth,angle=-90]{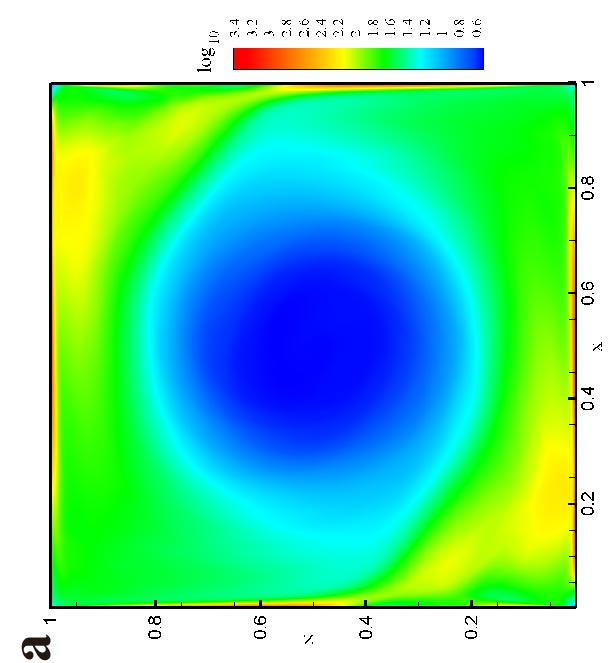}}
\subfigure[\label{subfig:Figure5b}]{\includegraphics[width =  0.21\textwidth,angle=-90]{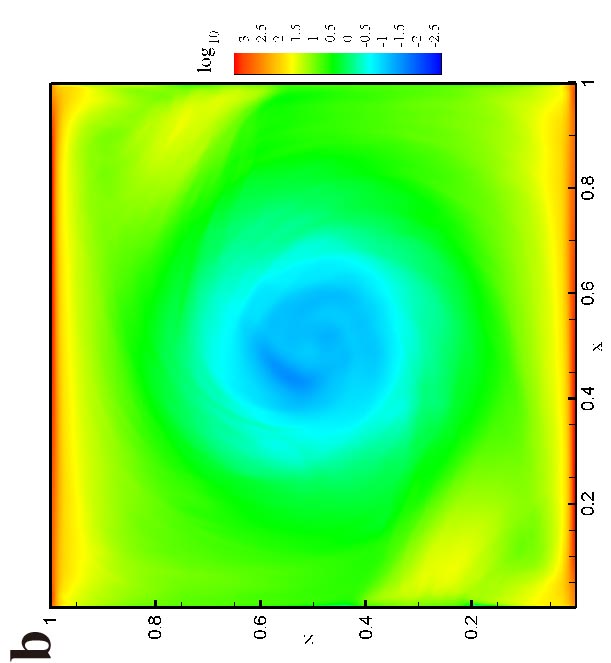}}
\subfigure[\label{subfig:Figure5c}]{\includegraphics[width =  0.21\textwidth,angle=-90]{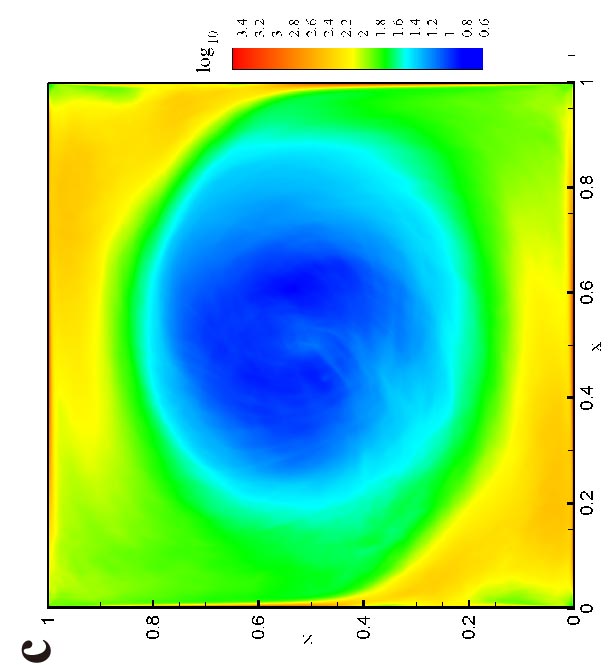}}
\subfigure[\label{subfig:Figure5d}]{\includegraphics[width =  0.21\textwidth,angle=-90]{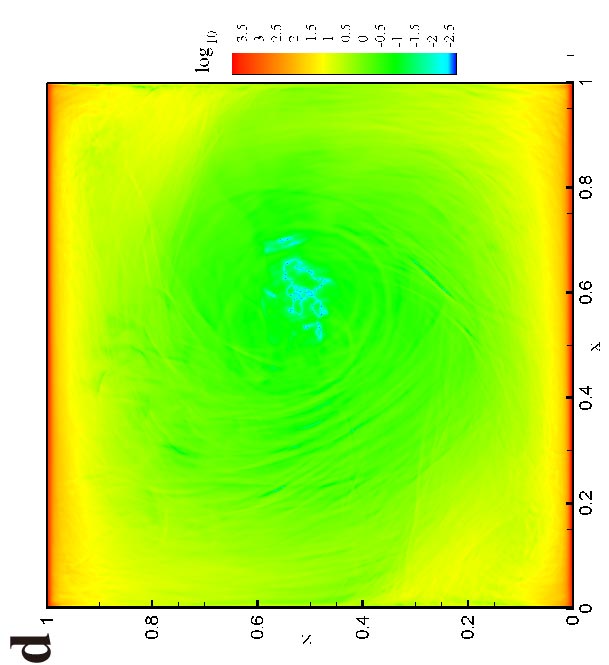}}
\end{center}
\caption{(Color online) Spatial average along $y$-axis of momentum dissipation $\varepsilon_{u}$ and thermal dissipation $\varepsilon_{\theta}$ for $Ra=1\times 10^8$ in (a)--(b) and for $Ra=1\times 10^9$ in (c)--(d), respectively.}
\end{figure*}

Fig. \ref{subfig:Figure5a} and \ref{subfig:Figure5c} show the distributions of momentum dissipation $\varepsilon_u$ for $Ra = 1\times10^8 $ and $Ra=1\times 10^9$, respectively, and the thermal dissipation $\varepsilon_{\theta}$ is shown in Fig. \ref{subfig:Figure5b} and \ref{subfig:Figure5d}. Note that the maximum of both $\varepsilon_u$ and $\varepsilon_{\theta}$ are extremely close to the bottom/top plates. On the other hand, the variation of $\varepsilon_{\theta}$ near the sidewalls is moderate, due to the adiabatic wall condition, while $\varepsilon_u$ exhibits a significant enhancement in the impacting region (i.e. left-down and right-up corner of Fig. \ref{subfig:Figure5a} and \ref{subfig:Figure5c}). As a contrast, both $\varepsilon_{\theta}$ and $\varepsilon_u$ are low in the bulk zone, corresponding to the core of LSC.

\subsection{Multilayer structure of velocity profiles \label{subsec:viscBL}}
\begin{figure}
\begin{center}
\includegraphics[width=0.5\textwidth]{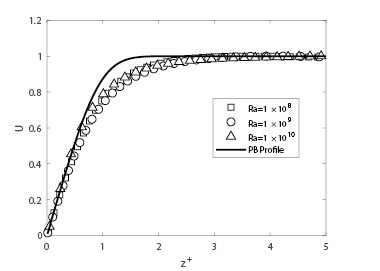}
\end{center}
\caption{The mean velocity profiles at the box center for the present study ($Ra=1\times 10^8 \sim 1\times 10^{10}$), with the  PBP profile (solid line). \label{fig:U-z}}
\end{figure}

The vertical MVPs at the center of the wind-shearing region ($x/L=0.5$) against the PBP-type profiles for different $Ra$s are shown in Fig. \ref{fig:U-z}. It is clearly seen that the MVPs dramatically differ from the PBP-type profile, especially near the conducting plates. The inset of Fig. \ref{fig:U-z} shows $U/U_{max}$ as a function of $z/\delta_v$ for different $Ra$s \cite{Sun2008JFM}. The deviation from the PBP-type profile was also reported in the literature (cf. \cite{duPuits2007PRL, Wei2013JFM}). We emphasize that the vertical acceleration of fluid in the ejecting region associated with the emitting plumes contributes to the deviation of the MVP from the PBP-type.

\begin{figure*}
\begin{minipage}[t]{0.5\linewidth}
\centering
\subfigure[\label{subfig:U-MVP}]{\includegraphics[width=.85\linewidth]{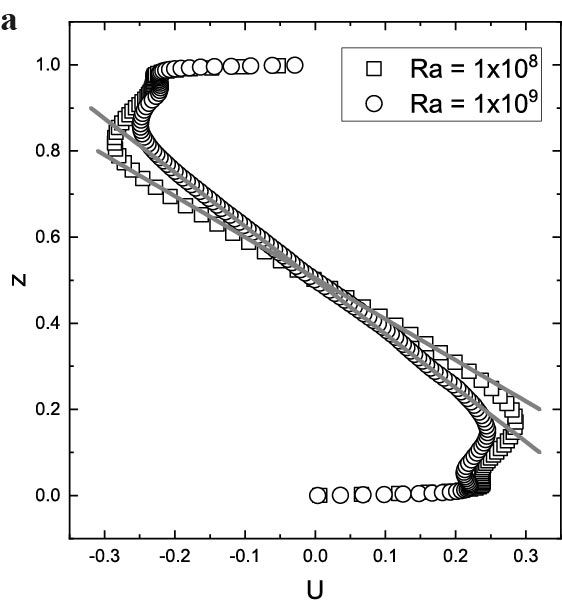}}
\end{minipage}
\begin{minipage}[t]{0.5\linewidth}
\centering
\subfigure[\label{subfig:W-MVP}]{\includegraphics[width=.85\linewidth]{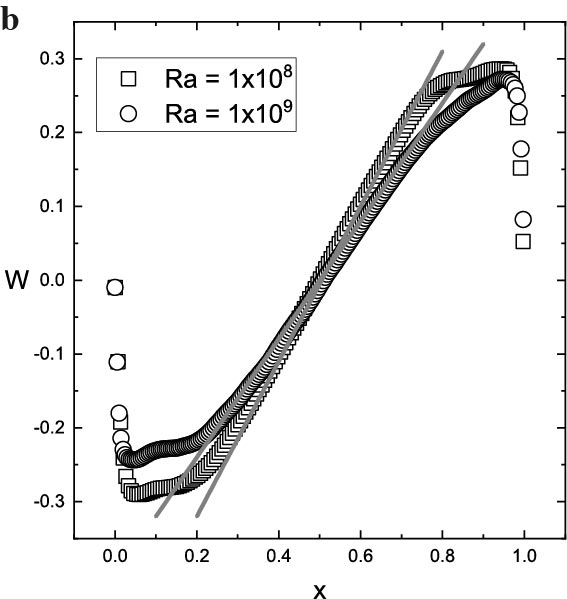}}
\end{minipage}
\caption{(a) Horizontal velocity profile $U(z)$ cut along the $z$ axis (at $x = 0.5$) and (b) vertical velocity profile $W(x)$ cut along the $x$ axis (at $z = 0.5$). \label{fig:UW-center}}
\end{figure*}

The mean horizontal velocity $U(z)$ at $x=L/2$, and the mean vertical velocity $W(x)$ at $z = L/2$ are shown in Fig. \ref{fig:UW-center}. A linear profile is found at the core of LSC for both $U(z)$ and $W(x)$, indicating a solid rotation of fluid in this region (for $0.25\leq z \leq0.75$ in Fig. \ref{subfig:U-MVP} and $0.25\leq x \leq0.75$ in Fig. \ref{subfig:W-MVP}). This has been experimentally observed in both rectangular \cite{Xia2003PRE} and cylindrical cell \cite{Qiu2001PRE}. The slim-box simulations show that the linear core of the LSC extending to around $0.4L$ for all $Ra$s, quite similar to that observed in a narrow 3D RBC cell by ref. \cite{Xia2003PRE} with $D/H \simeq 4$ and that in a cylindrical cell \cite{Qiu2001PRE}. We also notice that a hump of velocity appears at the border of the LSC from both $U$ and $W$ profiles in Fig. \ref{fig:UW-center}, which defines a characteristic LSC velocity, $U_{lc}$ and $W_{lc}$, respectively, distinct from peak velocity near the wall, which is denoted as $U_{nw}$ and $W_{nw}$, respectively.

Compared to the confined cell \cite{Xia2003PRE}, the present slim-box with periodic boundary has a higher $U_{lc}$ and a relatively large linear core. On the other hand, the $W_{lc}$ is the order of $W_{nw}$; see Fig. \ref{fig:UW-center}. Particularly, for horizontal velocity at $x/L=0.5$, $U_{lc}$ is greater than $U_{nw}$ for all $Ra$s, indicating that near-wall buoyancy is overwhelmed by the wind shearing. With increasing $Ra$, both $U_{lc}$ (and $W_{lc}$) and $U_{nw}$ (and $W_{nw}$) tend to decrease, but the linear core velocity has a more serious decrease than that near the wall.

\begin{figure*}
\begin{minipage}[t]{0.5\linewidth}
\centering
\subfigure[\label{subfig:lambdau-Ra}]{\includegraphics[width = .9\textwidth]{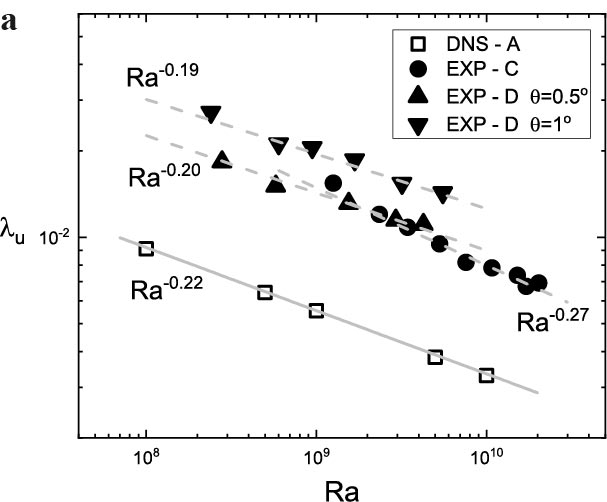}}\end{minipage}
\begin{minipage}[t]{0.5\linewidth}
\centering
\subfigure[\label{subfig:Re-Ra}]{\includegraphics[width = .9\textwidth]{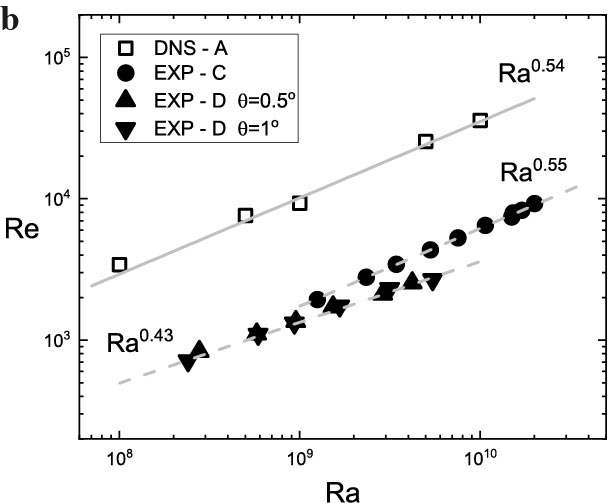}}
\end{minipage}
\caption{The Rayleigh number effects on the thickness of the viscous BL and the Reynolds number. (a) The $Ra$-dependence of the viscous boundary thickness normalized by the box height $H$. (b) The $Ra$-dependence of $Re$. \label{fig:Re-Ra}}
\end{figure*}

The thickness of the kinetic (viscous) BL as a critical quantity for turbulent BL is used here to characterize the present RBC-BL. Several definitions of the BL thickness have been given by \cite{Sun2008JFM}. The commonly used one is the wall distance obtained by extrapolation of the linear profile $U/U_{max} = z/\delta_v$ to reach $U_{max}$, denoted by $\lambda_u$. Fig. \ref{subfig:lambdau-Ra} shows that $\lambda_u/H$ is smaller than those measured in two experiments in confined cells \cite{Sun2008JFM, Wei2013JFM}. Note that a confined LSC tends to establish a more steady circulation in the cell. The experimental study with a tilted cylindrical cell shows that a larger tilt angle imposes stronger restriction on the azimuthal motion of the LSC so that it has less fluctuations perpendicular to the wind \cite{Wei2013JFM}. Moreover, the experiments of the slender rectangular cell also indicates that the confinement of the LSC tends to stabilize the large scale flow structure and lead to thinner viscous boundary layer \cite{Sun2008JFM}, as also seen in Fig. \ref{subfig:lambdau-Ra}. In the present study, the slim-box almost removes the azimuthal meandering of the LSC. The relatively small value of $\lambda_u$ for the slim-box flow is attributed to the absence of the depth confinement and thus the wall friction, which allows to develop a stronger LSC (with larger $U_{max}$) in a larger scale, leading to a thinner viscous BL in comparison with the confined cell. The present simulations present a scaling of $\lambda_u \sim Ra^{-0.27}$, similar to  the scaling of $-0.20$ from \cite{Wei2013JFM} , but different from that of \cite{Sun2008JFM}.


The mean velocity of LSC reaches a maximum $U_{max}$, relevant for determining the thicknesses of the velocity and thermal BLs. Here, we present the bulk Reynolds numbers defined by $Re=U_{max}H/\nu$ against $Ra$ obtained from the simulations and other two experiments in Fig. \ref{subfig:Re-Ra}. Due to a larger $U_{max}$ in the present simulations, $Re$s of our simulations are always larger than the experimental measurements. Again, the present results show a scaling ($Re\propto Ra^{0.54}$), close to that of \cite{Sun2008JFM} ($Re\propto Ra^{0.55}$). In comparison, the results of \cite{Wei2013JFM} show a smaller slope $Re\propto Ra^{0.43}$. This consistent scalings is also understood as the decease of the wall friction like the behavior of $\lambda_u$.

\begin{figure}
\includegraphics[width=\linewidth]{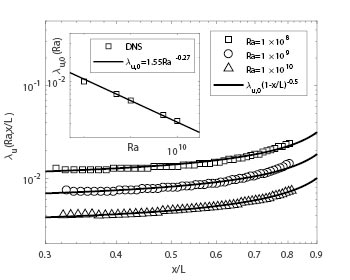}
\caption{The thickness of the viscous BL as a function $\lambda_u = \lambda_{u,0}(1-x/L)^{-0.5}$, represented by solid lines. \label{fig:lambdau-x}}
\end{figure}

The BL thickness over the whole wind shearing region is calculated (i.e. $0.3 < x < 0.8$) to illustrate the streamwise change of the viscous boundary layer. Fig. \ref{fig:lambdau-x} shows $\lambda_u(x)$ at three $Ra$s. A monotonous increase of the thickness can be describe in a power law, $\lambda_u = \lambda_{u,0}(1-x/L)^{-0.5}\propto r^{-0.5}$, where $r=1-x/L$ is the distance to the ejection corner of the cell. The inset of Fig. \ref{fig:lambdau-x} shows the variation of the coefficient $\lambda_{u,0}$ as a function of $Ra$, presenting a scaling law of $\lambda_{u,0} = 1\times Ra^{-0.27}$. Thus, the BL thickness seems to display a rather simple scaling: $\lambda_u(x)\approx  Ra^{-0.27}r^{-0.5}$ in the wind shearing region for all the simulations.

To quantify MVP in the wind-shearing region, we follow the SED theory \cite{She2017JFM} to employ the stress length as the similarity variable, which takes a multi-layer formula describing the structure in the $z$ direction normal to the wall. Neglecting the relatively small variation along the $x$ direction in the mean momentum equation (i.e. the incompressible boundary layer approximation), then one obtains the following balance equations:
\begin{eqnarray}\label{eq:MVWP0}
\nu\frac{\partial^2\overline{u}}{\partial z^2}-\frac{\partial\overline{u'w'}}{\partial z} = 0 \label{eq:MVP0};\\
\nu\frac{\partial^2\overline{w}}{\partial x^2}-\frac{\partial\overline{u'w'}}{\partial x} = 0 \label{eq:MWP0}.
\end{eqnarray}
Eq. (\ref{eq:MVP0}) holds for $\overline{u}$ in the wind-shearing region (near $x=L/2$), and Eq. (\ref{eq:MWP0}) for $\overline{w}$ near $z=H/2$. Integrating Eq. (\ref{eq:MVP0}) along $z$ and Eq. (\ref{eq:MWP0}) along $x$ yields
\begin{eqnarray}\label{eq:MVP11}
\frac{\partial\overline{u}^+}{\partial z^+} - \overline{u'w'}^+ = 1 \label{eq:MVP1};\\
\frac{\partial\overline{w}^+}{\partial x^+} - \overline{u'w'}^+ = 1 \label{eq:MWP1}.
\end{eqnarray}
Note that $u^+=u/u_{\tau}$ , $z^+=z/\delta_u$ and $(u'w')^+=(u'w')/u_{\tau}^2$ , where $u_{\tau}=\sqrt{\nu\partial{u}/\partial{z}|_{z=0}}$ and $\delta_u = \nu/u_{\tau}$. Introducing $S_u^+ = \frac{\partial\overline{u}^+}{\partial z^+}$ ($S_w^+ = \frac{\partial\overline{w}^+}{\partial x^+}$ ) and $W^+ = - \overline{u'w'}^+$ gives a normalized mean velocity equation as
\begin{eqnarray}
S_{u}^+(z^+) + W^+(z^+) = 1, \label{eq:MVP2}\\
S_{w}^+(x^+) + W^+(x^+) = 1. \label{eq:MWP2}
\end{eqnarray}
We emphasize that Eq. (\ref{eq:MVP2}) and (\ref{eq:MWP2}) hold when the relatively small pressure gradient effect in the wind shearing region is neglected. Introducing the stress length $\ell_{u,w}^+ = \sqrt{W^+}/S_{u,w}^+$, then, the solutions of Eq. (\ref{eq:MVP2}) and (\ref{eq:MWP2}) can be expressed as:
\begin{eqnarray}
\frac{\partial\overline{u}^+}{\partial z^+}\approx \frac{2}{1+\sqrt{1+4\ell_{u}^{+2}}}; \label{eq:Uequation} \\
\frac{\partial\overline{w}^+}{\partial x^+}\approx \frac{2}{1+\sqrt{1+4\ell_{w}^{+2}}}. \label{eq:Wequation}
\end{eqnarray}

The theory developed by She et al. \cite{She2017JFM} allows us to construct a two-layer similarity solution for $\ell_{u,w}^+ $, which is assumed to possess a dilation group invariance and to take the following analytic form
\begin{eqnarray}
\ell_{u}^+\approx \ell_{u0}^+(z^{+3/2})\left[1+\left(\frac{z^+}{z_{sub}^+}\right)^4\right]^{1/4}, \label{eq:lmu}\\
\ell_{w}^+\approx \ell_{w0}^+(x^{+3/2})\left[1+\left(\frac{x^+}{x_{sub}^+}\right)^4\right]^{1/4},\label{eq:lmw}
\end{eqnarray}
where $z^+_{sub}$ (or $x^+_{sub}$) is the thickness of the viscous sublayer near the bottom (or side) wall. Note that Eq. (\ref{eq:lmu}) and (\ref{eq:lmw}) present two asymptotic scalings --- for $z^+\ll z^+_{sub}$ (or $x^+\ll x^+_{sub}$), $\ell_{u}^+\sim z^{+3/2}$ (or $\ell_{w}^+\sim x^{+3/2}$); and for $z^+\gg z^+_{sub}$ (or $x^+\gg x^+_{sub}$), $\ell_{u}^+\sim z^{+5/2}$ (or $\ell_{w}^+\sim x^{+5/2}$), where the scaling exponent $3/2$ is derived in ref. \cite{She2017JFM}, while the exponent 5/2 is due to a transition of $\partial \overline{u}/\partial z$ (or $\partial \overline{w}/\partial x$) from $z^0$ (or $x^0$) in the sublayer to $z^{-1}$ (or $x^{-1}$) outside the sublayer, which is specific to the RBC.

\begin{figure*}
\begin{center}
\subfigure[\label{subfig:U_SED0}]{\includegraphics[width = .49\textwidth]{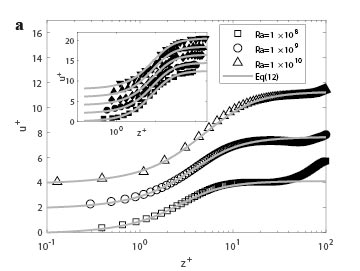}}
\subfigure[\label{subfig:W_SED0}]{\includegraphics[width = .49\textwidth]{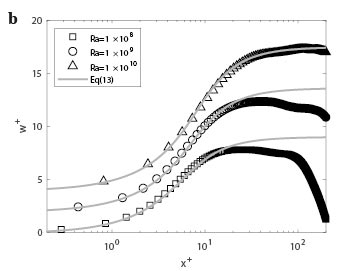}}
\subfigure[\label{subfig:U_rho_zsub}]{\includegraphics[width = .49\textwidth]{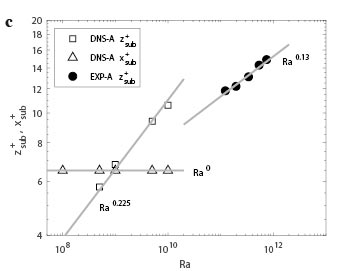}}
\subfigure[\label{subfig:lu0-lw0-Ra}]{\includegraphics[width = .49\textwidth]{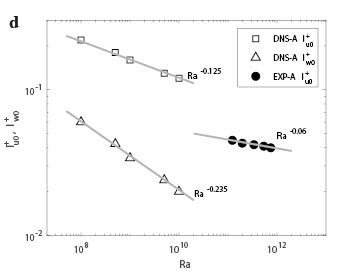}}
\end{center}
\caption{(a) Comparison of the multi-layer prediction for $U^+(z^+)$  (Eq. (\ref{eq:lmu}) plotted as solid line) and the DNS-A data (symbols). The velocity profiles from Eq. (\ref{eq:lmu}) for the EXP-A data are shown in the inset, where \ding{110}, \ding{108}, \ding{117}, \ding{115} and \ding{116}, represent $Ra=1.23\times 10^{11}$, $1.96\times 10^{11}$, $3.39\times 10^{11}$, $5.58\times 10^{11}$ and $7.48\times 10^{11}$, respectively. (b) Comparison of the multi-layer prediction for $W^+(z^+)$ from Eq. (\ref{eq:lmw}) and the DNS-A data. (c) The parameters, $z_{sub}^+$ and $x_{sub}^+$, as functions of $Ra$. (d) The parameters, $\ell_{u0}^+$ and $\ell^+_{w0}$, as functions of $Ra$.} \label{fig:U_SED0}
\end{figure*}

Fig. \ref{subfig:U_SED0} and \ref{subfig:W_SED0} show that the MVPs can be accurately described by the two-layer stress length formula (\ref{eq:lmu}) and (\ref{eq:lmw}). The two-layer structure extends its region up to $z^+\approx 30$ for $Ra = 10^{10}$. At lower $Ra$, the flow at $z^+ \sim 20$ is bulk-dominated, having a relatively thin boundary layer. The $Ra$-dependence of $\ell_0$ and $z^+_{sub}$ as in Fig. \ref{subfig:U_rho_zsub} shows that $\ell_0^+$ monotonously decreases like $\ell_{u0}^+=2.15Ra^{-0.125}$ in Eq. (\ref{eq:lmu}) and $\ell_{w0}^+ \approx 4.6\times Ra^{-0.235}$ in Eq. (\ref{eq:lmw}). The sublayer thickness $z_{sub}^+ \approx 0.062 \times Ra^{0.225}$ holds for two decades of $Ra$ ($1\times 10^8\le Ra\le 10^{10}$), whereas the convection in the cylindrical cell has a lower scaling, 0.13, as seen in Fig. \ref{subfig:U_rho_zsub}. This implies that the convection near the wall in a rectangular cell is stronger than that in a cylindrical cell, and intrigues to verify these scalings for different configurations over a wide rang of $Ra$. For the sublayer on the sidewalls, $x_{sub}^+ \approx 6.50$. The thickness of the sublayer is insensitive to $Ra$.

\subsection{Log-law of the thermal boundary layer \label{subsec:log-law-thermalBL}}
The logarithmic profile of the mean temperature distribution is one of the most remarkable results from recent turbulent RBC study. It is usually attributed to plume emission near the conducting plates \cite{vanderPoel2015PRL}.

According to the DNS, it is seen that thermal dissipation in the horizontal direction is almost ignorable in the ejecting region. Thus the temperature equation in the 2D form in ejecting region can be simplified as:
\begin{equation}\label{MTE}
- \frac{{\diff\overline{\theta} }}{{\diff z}} + \frac{{\overline{w\theta}}}{\kappa } = \left. {-\frac{{\diff\overline{\theta} }}{{\diff z}}} \right|_{z = 0}\equiv Nu.
\end{equation}
Denoting $S_\theta^* =-\mathrm{d}\overline{\theta}/(Nu\mathrm{d}z) =-\mathrm{d}\overline{\theta} /\mathrm{d}z^*$ and $W_\theta^* =\overline{w\theta}/(\kappa Nu)=\overline{w\theta}^*$ gives the normalized MTE
\begin{equation}\label{eq:scale-MTE}
S_\theta^*+W_\theta^*=1,
\end{equation}
with normalized vertical coordinate, $z^*=zNu$. One can describe similarities between the mean momentum equation for for RBC in three layers: (a) the near wall region (the conduction layer) is described by $ S_\theta^*\simeq 1$; (b) the region far away from the wall is dominated by $W_\theta^*\simeq 1$; and (c) the layer in between the above two layers. Note that $W_\theta^*$ represents the convective heat flux by normal velocity fluctuations from the wall. Thus, a thermal balance in form of the stress length similar to the momentum transport is obtained, and thus the distribution of the mean temperature is extracted.

We follow the SED theory \cite{She2017JFM} to quantify the mean temperature profile by postulating a similar (thermal) sublayer, buffer layer and log layer, with a temperature length $\ell _\theta$ as the similarity variable, which plays a similar role as the stress length. Specifically, $\ell _\theta$ displays a three-layer structure, that is, $\ell _\theta^*\propto\left( {z^*}\right)^{3/2}$ for $z^*<z_{sub}^*$; $ \ell _\theta^*\propto\left({z^*}\right)^{5/2}$ for $z_{sub}^*<z^*< z_{buf}^*$; and $\ell_\theta^*\propto z^*$ for $z^*>z_{buf}^*$, where the superscript $\ast$ denotes the variables normalization by $Nu/(2H)$. The first exponent ($3/2$) readily follows from a near-wall expansion with ignoring fluctuation magnitude, where $S_\theta^*\to 1$ and $W_\theta^*\to (z^*)^{3}$. The third layer has a linear scaling in $z$, corresponding to the log layer. The second layer is the buffer layer which, like the momentum buffer layer with a scaling different from the two above mentioned layers, is considered as the transition from a laminar flow near the wall to a fully turbulent state in the bulk. Its scaling ($5/2$) is obtained by invoking the integral-scale of temperature fluctuation, $\ell^*_{int}\equiv (W^*_\theta/S^*_\theta)^{3/4}/\epsilon_\theta^{1/4}$, which is proportional to $(z^*)^{9/4}$ near the wall, and a dissipation-production ratio for temperature fluctuation, $\Theta_\theta\equiv\epsilon_\theta/(S^*_\theta W^*_\theta)$, which is proportional to $z$ in the buffer layer by inspecting the DNS. Since $\ell^*_\theta = \ell^*_{int}\Theta_\theta^{1/4}$, then it follows $\ell^*_\theta\propto (z^*)^{5/2}$ in the buffer layer, as a semi-empirical result, to be derived more rigorously in the future. Finally, the SED theory postulates a generalized dilation symmetry that $\ell _\theta^*$ displays a generalized Lie-group invariance characterizing transition between the local scaling behaviors, yielding a composite solution of $\ell^*_{\theta}(z^*)$:
\begin{equation}
\ell^*_{\theta}(z^*)=\sqrt{W^*_\theta}/S^*_\theta = \ell^*_{\theta0}{z^*}^{3/2} \left[1+\left(\frac{z^*}{z^*_{sub}}\right)^{4}\right]^{\frac{1}{4}} \left[1+\left(\frac{z^*}{z^*_{buf}}\right)^{4}\right]^{-\frac{1.5}{4}}, \label{eq:dissip-leng}
\end{equation}
The stress length of temperature $\ell_\theta^*(z)$ is a three-layer function expressed as Eq. (\ref{eq:dissip-leng}). Transition of the scaling of $\ell^{*}_{\theta}$ from $3/2$ to $5/2$ occurs at $z^{*}_{sub}$. The next transition of scaling from $5/2$ to $1$ is present at $z^{*}_{buf}$.

Jointly solving (\ref{eq:scale-MTE}) and (\ref{eq:dissip-leng}) yields an analytical function of the MTP as:
\begin{equation} \label{eq:SED-theta}
\frac{1}{2}-\overline{\theta}(z^*) =  \int_0^{z^*} S_\theta \diff z'= \int_0^{z^*} \frac{{-1 + \sqrt {4\ell
			_\theta ^{*2} + 1} } }{2\ell _\theta ^{*2} } \diff z'.
\end{equation}
The first consequence of the solution is the logarithmic law of MTP. For $ z^ *  \gg z_{buf}^ *$, $ \ell _\theta ^ *   \approx \kappa
_\theta  z^ *\gg 1$, then $ S_\theta ^ *   \approx 1/\left( {\kappa_\theta  z^ *  } \right)$, a logarithmic MTP follows:
\begin{eqnarray}
\overline{\theta}  &\approx&  - \frac{1}{{\kappa _\theta  }}\ln z^ *   + B
=  - A\ln z^* + B\label{eq:logTheta} \\
1/A &=&\kappa _\theta=\ell^*_{\theta0}{z^*_{buf}}^{3/2}/z^*_{sub} \label{eq:kappa}
\end{eqnarray}

\begin{figure}
\begin{center}
\includegraphics[width = 0.49\textwidth]{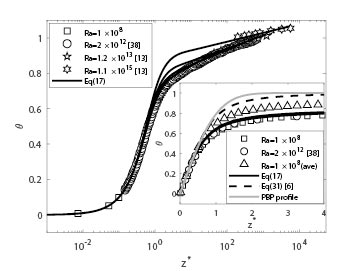}
\end{center}
\caption{The MTP of three data sets: the present DNS, the DNS from ref. \cite{Stevens2011JFM}, and the experiment from Ahlers et al. \cite{Ahlers2012PRL}. The solid lines are Eq. (\ref{eq:SED-theta}). The triangles are the MTP averaged along the horizontal ($x$) direction. The dashed curve is Eq. (31) from Shishkina et al. \cite{Shishkina2015PRL}, and the gray curve is PBP profile. \label{fig:SEDT-prof}}
\end{figure}

\begin{figure}
	\begin{center}
	\includegraphics[width =0.9\linewidth]{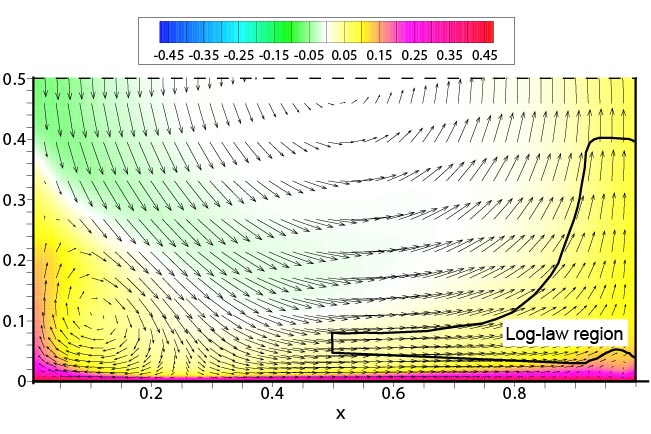}
	\end{center}
\caption{(Color online) The log-law region of thermal BL described by Eq. (\ref{eq:logTheta}) at $Ra=1\times10^8$, marked by solid line. The pseudo-color denotes temperature. Cold color represents low temperature, and hot color high temperature. The arrows are the velocity vectors. \label{fig:T-log-region}}
\end{figure}

Fig. \ref{fig:SEDT-prof} shows the comparison between the analytical solutions at a fixed location from the side wall ($r=0.045$) for $Ra$ covering seven decades, from moderate ($Ra = 1\times10^8$) in DNS to the extremely high $Ra$ experiments ($Ra=1\times 10^{15}$) \cite{Ahlers2012PRL}. The mean temperature profile (MTP) in the ejecting region clearly presents a log-law covering at least one decade in $z$ ($0.04 \leq z/H \leq 0.4$). Close inspection of Fig. \ref{fig:TP-field} and \ref{fig:T-log-region} shows that the range of log-layer coincides with the intensive plume emission, indicating the relation between the two phenomena.

\begin{figure*}
\begin{minipage}[t]{0.5\linewidth}
\centering
\subfigure[\label{subfig:ell-r}]{\includegraphics[width=.85\textwidth]{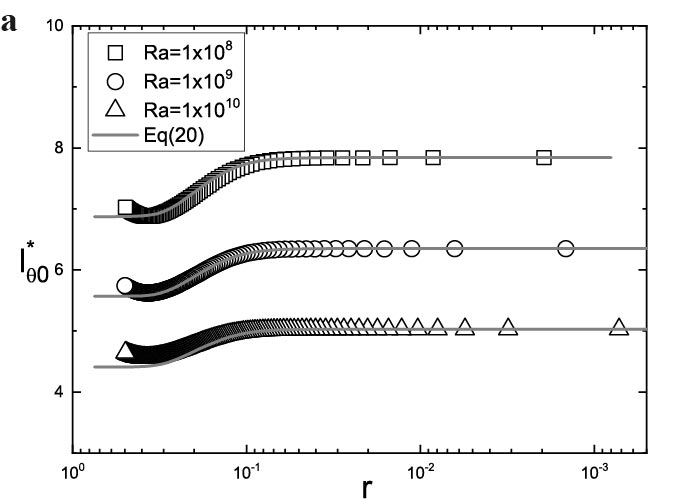}}
\end{minipage}
\begin{minipage}[t]{0.5\linewidth}
\centering
\subfigure[\label{subfig:zbuf-r}]{\includegraphics[width=.85\textwidth]{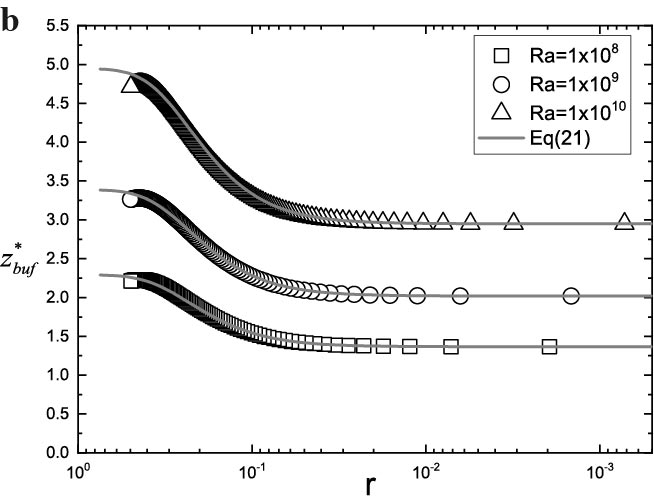}}
\end{minipage}
\caption{(a) $\ell^*_{\theta 0}(r)$ for $Ra=1\times10^8$, $Ra=1\times10^9$  and $1\times10^{10}$, respectively. The solid lines are the multilayer function of Eq. (\ref{eq:ell0_x}). (b) $z^*_{buf}(r)$. The solid lines are the multilayer function of Eq. (\ref{eq:zbuf_x}).\label{fig:ell-zbuf-r}}
\end{figure*}

We calculated the theoretical model, Eq. (31) for $Pr \gtrsim 1$ from ref. \cite{Shishkina2015PRL}, with parameter $c=1$; see the inset of Fig. \ref{fig:SEDT-prof}. The systematic deviations of the MTP ($Pr<1$) from the equation and the PBP profile predictions are observed. The time average temperature profile is located in the ejecting region, where the wind is turning its direction and the near wall flow is no longer the typical turbulent boundary layer. Thus we calculated the MTP averaged along the horizontal ($x$) direction, which represents the characteristics of the boundary layer in the wind shearing region \cite{Shishkina2015PRL}. The inset of Fig. \ref{fig:SEDT-prof} shows that, the space-averaged MTP is remarkably higher than the local MTP in the ejecting region, and much closer to the BL equation. Though the MTP in the wind-shearing region with small $Pr$ has a similar trend as the case with large $Pr$, neither the present MTP nor that from Stevens et al. \cite{Stevens2011JFM} ($Pr=0.7$) can be described by the equation for $Pr \gtrsim 1$. However, an improved thermal boundary layer equation is capable of describing the flow with low Prandtl number down to $0.01$ \cite{Shishkina2017PRF}.

Three parameters, $\ell^*_{\theta 0}$, $z^*_{sub}$ and $z^*_{buf}$  are determined by the profile of $l^{*}_{\theta}(z^*)$. Since thermal dissipation dominates the boundary layer near the side wall ($r<0.1$ for DNS-A), the heat transport from convection is neglected in this region. Measurement of $z^*_{sub}$ indicates that $z^*_{sub} \simeq 0.375$. Two other parameters, $\ell^*_{\theta 0}$ and $z^*_{buf}$, are found $r$-dependent. The $r$-dependence of $\ell^*_{\theta 0}$ and $z^*_{buf}$ are expressed in form of two-layer structures, expressed as the functions of $r=(1-x/L)$:
\begin{eqnarray}
\ell^*_{\theta 0} = \ell^*_{\theta 0,a} \left[1+\left(\frac{r}{r_{b}}\right)^{4}\right]^{-\frac{0.15}{4}} \left[1+\left(\frac{r}{r_{c}}\right)^4\right]^{\frac{0.15}{4}}; \label{eq:ell0_x} \\
z^*_{buf}= z^*_{buf,a} \left[1+\left(\frac{r}{r_{b}}\right)^{2}\right]^{\frac{0.5}{2}} \left[1+\left(\frac{r}{r_{c}}\right)^4\right]^{-\frac{0.5}{4}}. \label{eq:zbuf_x}
\end{eqnarray}
Near the side wall ($r\ll r_{b}$), $\ell^*_{\theta 0}$ and $z^*_{buf}$ are constants, to be denoted as $\ell^*_{\theta 0,a}$ and $z^*_{buf,a}$. The coefficient, $\ell^*_{\theta 0,a}$, may be associated with the size of thermal plumes, which decreases for increasing $Ra$, to be determined below. In the wind-shearing region ($r\gg r_{b}$), we have a scaling: $\ell^*_{\theta 0}\sim r^{-0.15}$, which drops with increasing $r$; see Fig. \ref{subfig:ell-r}. On the other hand, the thickness of buffer layer $z^*_{buf}$ as shown in Fig. \ref{subfig:zbuf-r}, increases in $r$, indicating that the thermal boundary layer becomes thinner, going downstream with the wind (for increasing $x$ or decreasing $r$).
 
\begin{figure*}
\begin{minipage}[t]{0.49\textwidth}
\centering
\subfigure[\label{subfig:ellstar-Ra}]{\includegraphics[width=.95\textwidth]{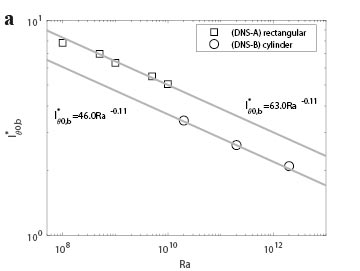}}
\end{minipage}
\begin{minipage}[t]{0.49\textwidth}
\centering
\subfigure[\label{subfig:zstarbuf-Ra}]{\includegraphics[width=.95\textwidth]{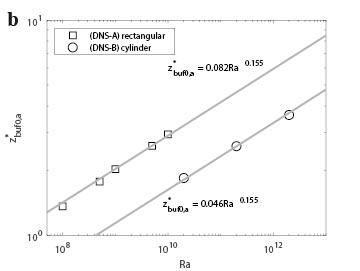}}
\end{minipage}
\caption{ The $Ra$-dependence of $\ell^*_{\theta 0,a}$ (a) and $z^*_{buf 0,a}$ (b). \label{fig:zbufa0-ella0-Ra}}
\end{figure*}

After comparing the data sets, we find that $r_{b} \simeq 0.125$ and $r_{c} \simeq 0.35$ for the slim-box (for $DNS-A$), but $\simeq 0.0075$ and $0.35$ for the cylindrical cell ($DNS-B$) --- they are likely independent of $Ra$. On the other hand, the coefficients, $\ell^*_{\theta 0,a}$ and $z^*_{buf 0,a}$, depend on $Ra$; see Fig. \ref{fig:zbufa0-ella0-Ra}. Specifically, $z^*_{buf 0,a} \approx 0.082 Ra^{0.155}$ for the slim-box simulations ($DNS-A$), but $0.046 Ra^{0.155}$ for the cylindrical cell ($DNS-B$). The same scaling of $0.155$ for $Ra$ indicates the similar behavior of the heat transport near the side walls. Moreover, $\ell^*_{\theta 0,a} \simeq 63.0 Ra^{-0.11}$ for the slim-box ($DNS-A$), but $46.0 Ra^{-0.11}$ for the cylindrical cell, also with the same scaling of $-0.11$ (see the inset of Fig. \ref{fig:zbufa0-ella0-Ra}). Since $z^*_{sub} \simeq 0.375$ is independent of $Ra$ and geometry, matching condition for $1/A \simeq [\ell^*_{\theta 0} {z^*_{buf}}^{3/2} / z^*_{sub}]$ yields a scaling for the coefficient of the log law slope, $A \sim Ra^{0.1225}$ near the side wall, in good agreement with experimental result of Ahlers et al. \cite{Ahlers2014JFM}: $A \sim Ra^{0.123}$. 

\begin{figure}
\centering
\includegraphics[width = 0.97\linewidth]{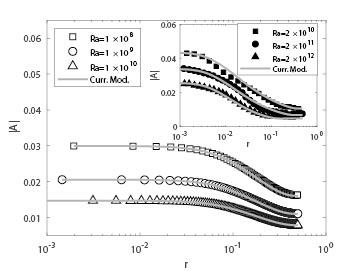}
\caption{The parameter $A$ of the log-law function varies with the longitudinal coordinate $x$ for different $Ra$s in DNS-A. The inset is the result of DNS-B. The solid lines represent $A \simeq z^*_{sub}/\left(\ell^*_{\theta 0}{z^*_{buf}}^{3/2}\right)$. \label{fig:A-x}}
\end{figure}

The coefficient in the log-law $A = 1/\kappa_{\theta}$ as the function of $x$ was first discussed in ref. \cite{Grossmann2012PoF}, suggesting an analytic form, $|A|=|A_1|/\sqrt{\left[4x(1-x)\right]}$ for the cylindrical cell. Note that $A(x)$ holds a $-1/2$--scaling (i.e. $A\sim A_0/\sqrt{x}$ for $x\to 0$). Lately, Ahlers et al. \cite{Ahlers2012PRL} claimed that the scaling was $-0.65$, due to an elliptical path (rather than a circular path) of LSC. In the present study, for $z^*\gg z^*_{buf}$, Eq. (\ref{eq:dissip-leng}) can be rewritten as $\ell^*_{\theta} = [\ell^*_{\theta 0} {z^*_{buf}}^{3/2} / z^*_{sub}] z^*$, i.e. $\kappa_{\theta} =1/A \simeq [\ell^*_{\theta 0}{z^*_{buf}}^{3/2}/z^*_{sub}]$. Based on the aforementioned analysis on $\ell^*_{\theta}$ and $z^*_{buf}$ (i.e. $\ell^*_{\theta 0} \sim r^{-0.15}$, $z^*_{buf}\sim r^{0.5}$, and $z^*_{sub}=\mathrm{const.}$), we have $A\sim r^{-0.60}$, in agreement with the recent result of Grossmann et al. \cite{Grossmann2012PoF}. Fig. \ref{fig:A-x} is the comparison between our predicted coefficient $A(r)$ from DNS at different $Ra$s, which is superior to the fitting function of \cite{Grossmann2012PoF}; the latter is only valid in a restricted domain ($0.02<r<0.2$ for $Ra=2\times10^{12}$ from ref. \cite{Ahlers2012PRL}). Thus, the current model gives a unified description of $A$ valid over the wider flow domain, from the side wall ($r=0$) to the wind shearing region, reflecting the behaviors in different regions.

\begin{figure}
\begin{center}
\includegraphics[width=.9\linewidth]{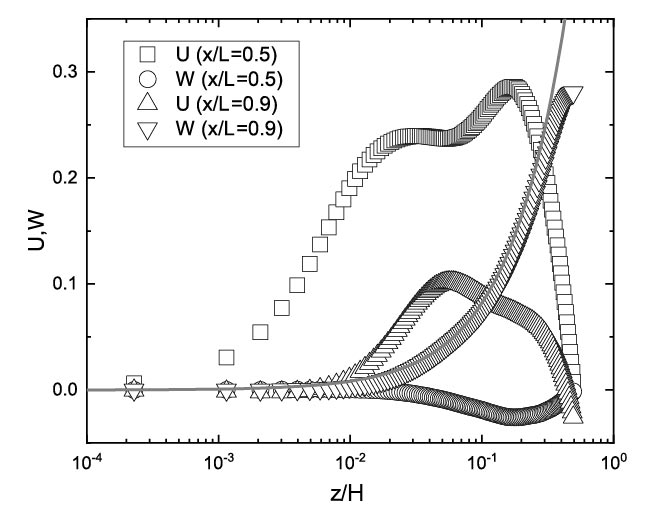}
\end{center}
\caption{The horizontal $U(z)$ and vertical velocity $W(z)$ profiles in ejecting region at $x/L=0.9$ and the wind-shearing region at $x/L=0.5$ for the case of $Ra = 1\times 10^8$. The solid line is the fitting of $W/U_0=0.7(z/H)$. \label{fig:UW-region}}
\end{figure}

Furthermore, to investigate the dynamical mechanism of the log-law of temperature, we calculated the profiles of the horizontal and vertical velocity in the ejecting and wind-shearing regions, as presented in Fig. \ref{fig:UW-region}. Note that there is no logarithmic region for $U(z)$ in the ejecting region. Thus we believe that the log-layer of temperature has an origin free from the log-law of the mean velocity; in other words, the simple Reynolds analogy does not hold here. It is inspiring to make a comparison between velocities in the wind-shearing region ($0.25 \lesssim x/L \lesssim 0.75$) and those in the ejecting region ($x/L \gtrsim 0.75$). The latter is governed by strong APG due to the confinement of the sidewalls, which transforms horizontal momentum (in the wind-shearing region) into upward (vertical) momentum in the ejecting region. Since the ejecting region overlaps with the log temperature region, we believe that the vertical momentum plays the main role in establishing the log-law of temperature. In addition, the vertical velocity is approximately proportional to the wall distance, $W/U_0 = 0.7(z/H)$ at $0 \leq z/H \leq 0.4$ in the ejecting region, as shown in Fig. \ref{fig:UW-region}. The theoretical work taking into account this observation for a complete explanation of the log law of mean temperature will be reported elsewhere.

\subsection{Rayleigh number effect on heat transport}
\begin{figure}
\begin{center}
\includegraphics[width = .85\linewidth]{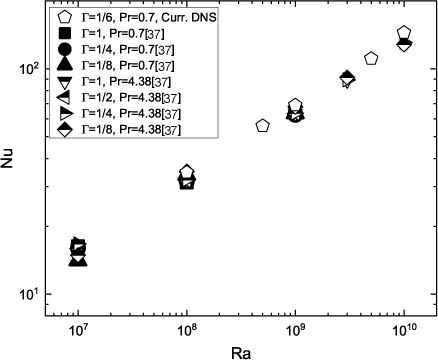}
\end{center}
\caption{Compensated $Nu$ as a function of $Ra$. \label{fig:Nu-Ra}}
\end{figure}

One of the main issues in the study of RBC is to determine the dimensionless heat transfer coefficient, the Nusselt number $Nu$. The heat flux can be calculated by $Nu(z) = \left\langle  w \theta / \kappa - \partial\theta/\partial{z} \right\rangle_{x,y,t}$, where $\left\langle \cdot \right\rangle_{x,y,t}$ represents averaging over the horizontal plane and sufficiently long time \cite{Chong2016JFM}. A comparison between $Nu$ of present simulation with previous data is presented in Fig. \ref{fig:Nu-Ra}.  Note that the DNS by Kaczorowski et al. \cite{Kaczorowski2014JFM} examined various aspect ratios (varying from $\Gamma=1/8$ to $1$) and Prandtl numbers ($Pr=0.7$ and $4.3$). The comparison with other data at the same $Pr$ (i.e. $Pr=0.7$) shows that our slim-box simulations are mostly valid, but our $Nu$ is slightly larger than that of \cite{Kaczorowski2014JFM}, consistent with the fact that the slim-box tends to enhance the LSC and thus a more intensive heat transport.

\begin{figure*}
\subfigure[\label{fig:locNubuttom}]{\includegraphics[width = .52\textwidth]{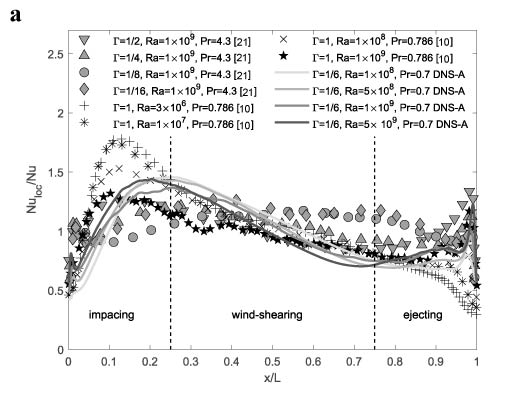}}
\subfigure[\label{fig:locNuhalf}]{\includegraphics[width = .47\textwidth]{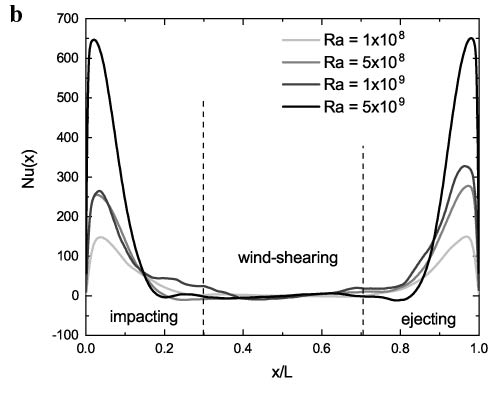}}
\caption{(a) Local heat flux $Nu_{loc}/Nu$ (normalized with global Nusselt number, $Nu$) at the bottom plate (i.e. $z=0$) varying with $x$ \cite{Wagner2012JFM, Huang2013PRL}. (b) $Nu_{loc}$ at the half height of the convection box ($z=H/2$) varying with $x$.}
\end{figure*}

We now examine the contribution to the total heat transport from different flow region, from $Nu_{loc}(x,z)=\left\langle w\theta/\kappa-\nabla\theta\right\rangle_{y,t}$, where $\left\langle \cdot \right\rangle_{y,t}$ denotes the averaging over the depth direction and sufficiently long time. Fig. \ref{fig:locNubuttom} shows $Nu_{loc}(x,0)/Nu$ at the bottom plate (i.e. $z=0$), which is dominated by thermal diffusion $Nu_{loc}(x,0) \approx -\nabla\theta(x,0)$. Note that the maximum heat flux appears at $x = 0.25$, where the thickness of the thermal BL is the thinnest, as shown in Fig. \ref{fig:locNubuttom}, corresponding to where the cold plumes impinge on the heating plate. The total heat flux in the range of $0 \leq x \leq 0.25$ corresponds to that by the corner roll.

In the center region ($0.25\lesssim x/L \lesssim0.75$) \cite{Xia2003PRE, Wagner2012JFM}, our computation shows a linearly decreasing heat flux with increasing $x$, corresponding to an increase of the thermal BL thickness, consistent with other data. However, this decreasing trend is weakened at higher $Ra$ and the flow becomes more homogeneous horizontally, which is sound. Fig. \ref{fig:locNuhalf} shows the local heat flux $Nu_{loc}$ at the half height of the box ($z/H=1/2$). For $Ra=10^8$, $Nu_{loc}$ is more symmetrical with respect to $z=1/2$ than $z=0$, which is due to the heat convection by LSC.
The three flow regions, i.e. impacting ($0 \leq x/L \leq 0.25$), wind-shearing ($0.25 \leq x/L \leq 0.75$) and ejecting ($0.75 \leq x/L \leq 1$) regions are rather distinct, as shown in Fig. \ref{fig:locNubuttom}, which is also clearly observed for $Ra=10^9$. Convective heat transfer is predominant in heat flux near the centerline $z/H=1/2$. We find that there is a symmetry breaking in the vertical velocity $W(x)$ near the centerline $z=1/2H$, the maximum magnitude of $W$ is $0.271$ on the right side, greater than $-0.243$ on the left side, leading to the higher local heat flux at $x = 0$ side.

\section{Concluding Remarks}
We performed the 3D DNS of RBC at $Pr=0.7$ and $Ra=1\times10^8 \sim 1\times10^{10}$ for a slim-box (the ratio of length, depth and height is $1:1/6:1$), with periodic boundary condition in the depth ($y$) direction. Two major features occur in this slim-box RBC: the LSC is steadily confined parallel to $xz$-plane, and the wall friction on this plane is absent, resulting in a higher heat flux and stronger LSC. Three flow regions (i.e. impacting, wind-shearing, and ejecting) were studied in detail. \\The non-Blassius velocity profile under the influence of strong adverse pressure gradient can be described by a multi-layer stress length function, following the symmetry-based theory of canonical wall turbulence \cite{She2017JFM}. The mean temperature profile can also be satisfactorily described by a multi-layer structure of a thermal dissipation stress length function, which yields an analytic description of the log-law coefficient $A$ for a range of $x$ and $Ra$. A few of parameters, like $\ell_{u0}$, $\ell_{w0}$, $\ell_{\theta 0}$, and $z_{buf}$, in the stress functions for viscous and thermal boundary layers depend on $Ra$, $Pr$, or even the geometry of the convection cell. Parameterization of the multilayer functions of the stress lengths for different configurations can be fulfilled by investigating more experiments and numerical simulations. Moreover, with knowledge of the symmetry in each layer, the multilayer functions can be applied to interpret and predict the convection flow at extreme conditions, for example, high $Ra$ or high/low $Pr$. \\Local heat transport was also discussed. The Nusselt number scaling and local heat flux of the present simulations are consistent with previous experiments \cite{Huang2013PRL} and numerical simulations \cite{Kaczorowski2014JFM} in the confined rectangular cell. Thus, we conclude that the present slim-box RBC is an ideal system for studying in-box kinetic and thermal structures, and space-time correlations \cite{He2014AMS}, in confined turbulent convection.


\vskip 1.5\kh

 \n{\bf \small References}

\vskip5mm

\footnotesize
\def\hang{\hangindent\parindent}
\def\textindent#1{\indent\llap{#1\enspace}\ignorespaces}
\def\re{\par\hang\textindent}
\parindent=15pt


\begin{acknowledgements}
The project supported by the National Natural Science Foundation of China (11452002, 11521091, and 11372362) and MOST(China) 973 project 2009CB724100.
\end{acknowledgements}


%

\end{document}